\documentclass[11pt]{article}
\pdfoutput=1
\usepackage[utf8]{inputenc}
\usepackage{dsfont}
\usepackage{diagbox}
\usepackage{amsfonts}
\usepackage{mathtools}
\allowdisplaybreaks[4]        
\usepackage{amssymb}
\usepackage{euscript}     
\usepackage{braket}
\usepackage{starfont}
\usepackage{color,soul}         
\usepackage{braket}
\usepackage{tensor}        
\usepackage{amsthm}
\usepackage{graphicx}
\usepackage{slashed}
\usepackage{leftidx}
\usepackage{subfigure}
\usepackage{bbm}
\usepackage{indentfirst}

\definecolor{outerspace}{rgb}{0.25, 0.29, 0.3}
\definecolor{scarlet}{rgb}{1.0, 0.13, 0.0}
\usepackage[header,title,page,titletoc]{appendix}  
\definecolor{princetonorange}{rgb}{1.0, 0.56, 0.0}
\definecolor{WildStrawberry}{rgb}{1.0, 0.26, 0.64}
\definecolor{rossocorsa}{rgb}{0.83, 0.0, 0.0}
\definecolor{navyblue}{rgb}{0.0, 0.0, 0.5}
\usepackage[numbers,sort&compress]{natbib}  
\usepackage{float}
\usepackage[paper=a4paper,margin=1in]{geometry}
\usepackage{titlesec}

\pdfoutput=1
\usepackage{amsmath}
\usepackage{color}
\usepackage{amsfonts}
\usepackage{amssymb}
\usepackage{graphicx}
\usepackage{geometry}
\usepackage{amssymb,epsfig,subfigure}
\usepackage{amssymb}
\usepackage{hyperref}
\usepackage{comment}
\usepackage[font=footnotesize]{caption}
\usepackage[T1]{fontenc}
\usepackage[utf8]{inputenc}
\usepackage{latexsym}

\usepackage{cancel}

\usepackage[numbers,sort&compress]{natbib}  
\usepackage{float}

\usepackage{dsfont}
\usepackage{diagbox}

\usepackage{mathtools}
\allowdisplaybreaks[4]

\usepackage{euscript}     
\usepackage{braket}
\usepackage{starfont}
\usepackage{color,soul}         
\usepackage{tensor}        
\usepackage{amsthm}

\usepackage{slashed}
\usepackage{leftidx}
\usepackage{bbm}

\makeatletter
\renewcommand\section{\@startsection {section}{1}{\z@}%
                                 {-3.5ex \@plus -1ex \@minus -.2ex}
                                   {2.3ex \@plus.2ex}%
                                   {\normalfont\large\bfseries}}
\renewcommand\subsection{\@startsection{subsection}{2}{\z@}%
                                   {-3.25ex\@plus -1ex \@minus -.2ex}%
                                     {1.5ex \@plus .2ex}%
                                     {\normalfont\bfseries}}
\renewcommand\subsubsection{\@startsection{subsubsection}{3}{\z@}%
                                   {-3.25ex\@plus -1ex \@minus -.2ex}%
                                     {1.5ex \@plus .2ex}%
                                     {\normalfont\itshape}}
\makeatother

\def\pplogo{\vbox{\kern-\headheight\kern -29pt
\halign{##&##\hfil\cr&{\ppnumber}\cr\rule{0pt}{2.5ex}&\ppdate\cr}}}
\makeatletter
\def\ps@firstpage{\ps@empty \def\@oddhead{\hss\pplogo}%
  \let\@evenhead\@oddhead 
}
\def\maketitle{\par
 \begingroup
 \def\thefootnote{\fnsymbol{footnote}}
 \def\@makefnmark{\hbox{$^{\@thefnmark}$\hss}}
 \if@twocolumn
 \twocolumn[\@maketitle]
 \else \newpage
 \global\@topnum\z@ \@maketitle \fi\thispagestyle{firstpage}\@thanks
 \endgroup
 \setcounter{footnote}{0}
 \let\maketitle\relax
 \let\@maketitle\relax
 \gdef\@thanks{}\gdef\@author{}\gdef\@title{}\let\thanks\relax}
\makeatother

\numberwithin{equation}{section}

\newcommand\eea{\end{eqnarray}}
\newcommand\bea{\begin{eqnarray}}

\def\beq{\begin{equation}}
\def\eeq{\end{equation}}

\newcommand{\be}{\begin{equation}}
\newcommand{\ee}{\end{equation}}
\newcommand{\ba}{\begin{align}}
\newcommand{\ea}{\end{align}}
\newcommand{\bg}{\begin{gather}}
\newcommand{\eg}{\end{gather}}
\newcommand{\bseq}{\begin{subequations}}
\newcommand{\eseq}{\end{subequations}}

\renewcommand{\ln}{\mathop{\rm ln}\nolimits}

\newcommand{\tr}{{\rm tr}}

\usepackage{hyperref}
\hypersetup{
    colorlinks,
    citecolor=rossocorsa,
    filecolor=navyblue,
    linkcolor=navyblue,
    urlcolor=navyblue
}

\begin{document} 

\begin{titlepage}

\begin{center}

\phantom{ }
\vspace{3cm}

{\bf \Large{Entropic order parameters in weakly coupled gauge theories}}
\vskip 0.5cm
Horacio Casini${}^{*}$,  Javier M. Mag\'an${}^{\ddagger}$, Pedro J. Mart\' inez${}^{\mathsection}$
\vskip 0.05in
\small{${}^{*}$ ${}^{\mathsection}$ \textit{Instituto Balseiro, Centro At\'omico Bariloche}}
\vskip -.4cm
\small{\textit{ 8400-S.C. de Bariloche, R\'io Negro, Argentina}}
\vskip -.10cm
\small{${}^{\ddagger}$ \textit{David Rittenhouse Laboratory, University of Pennsylvania}}
\vskip -.4cm
\small{\textit{ 209 S.33rd Street, Philadelphia, PA 19104, USA}}
\vskip -.10cm
\small{${}^{\ddagger}$ \textit{Theoretische Natuurkunde, Vrije Universiteit Brussel (VUB) and The International Solvay Institutes}}
\vskip -.4cm
\small{\textit{ Pleinlaan 2, 1050 Brussels, Belgium}}

\begin{abstract}

The entropic order parameters measure in a universal geometric way the statistics of non-local operators responsible for generalized symmetries. In this article, we compute entropic order parameters in weakly coupled gauge theories. To perform this computation, the natural route of evaluating expectation values of physical (smeared) non-local operators is prevented by known difficulties in constructing suitable smeared Wilson loops. We circumvent this problem by studying the smeared non-local class operators in the enlarged non-gauge invariant Hilbert space. This provides a generic approach for smeared operators in gauge theories and explicit formulas at weak coupling. In this approach, the Wilson and 't Hooft loops are labeled by the full weight and co-weight lattices respectively. We study generic Lie groups and discuss couplings with matter fields. Smeared magnetic operators, as opposed to the usual infinitely thin ones, have expectation values that approach one at weak coupling.  The corresponding entropic order parameter saturates to its maximum topological value, except for an exponentially small correction, which we compute. On the other hand, smeared 't Hooft loops and their entropic disorder parameter are exponentially small. We verify that both behaviors match the certainty relation for the relative entropies. In particular, we find upper and lower bounds (that differ by a factor of 2) for the exact coefficient of the linear perimeter law for thin loops at weak coupling. This coefficient is unphysical/non-universal for line operators.  We end with some comments regarding the RG flows of entropic parameters through perturbative beta functions.

\end{abstract}
\end{center}

\small{\vspace{3 cm}\noindent${}^{\text{\text{*}}}$casini@cab.cnea.gov.ar\\
${}^{\dagger}$magan@sas.upenn.edu\\
${}^{\mathsection}$pedro.martinez@cab.cnea.gov.ar}

\end{titlepage}

\setcounter{tocdepth}{2}

{\parskip = .4\baselineskip \tableofcontents}
\newpage

\section{Introduction}

A new perspective on generalized symmetries in quantum field theory (QFT) emerges by analyzing basic properties of the way observables are attached to spacetime regions \cite{Casini:2020rgj}. Given a region $R$ with associated algebra $\mathcal{A}(R)$,  generated by local degrees of freedom,  causality takes the  form
\be \label{caus}
\mathcal{A}(R)\subset \mathcal{A}(R')'\;,
\ee
where $R'$ refers to the causal complement of the region $R$, i.e., the points spatially separated from $R$, and $\mathcal{A}'$ is the commutant of $\mathcal{A}$, i.e., the set of operators that commute with $\mathcal{A}$. Eq. (\ref{caus}) just expresses the usual commutativity of spatially separated observables.   The proposal of \cite{Casini:2020rgj} is that the non-saturation of~(\ref{caus}) implies the existence of a certain generalized symmetry. See \cite{Review} for a brief review and \cite{Casini:2019kex} for an extensive treatment of the case of global symmetries. This approach is inspired in the algebraic approach to global symmetries, the so-called DHR approach \cite{Doplicher:1969tk,Doplicher:1969kp,Doplicher:1971wk,Doplicher:1973at,Doplicher:1990pn}. But while in this latter the starting point is given by the study of superselection sectors and the associated endomorphisms of the observable algebra,  \cite{Casini:2020rgj}  focuses on the non-saturation of causality as the origin of symmetry.

The nature of such symmetry depends on the topology of the regions in which the non-saturation occurs. As we review below, pure gauge theories display such violations for ring-like regions with non-trivial homotopy group $\pi_{1}$, and dually for regions with non-trivial $\pi_{d-3}$, for theories in $d$ dimensions. In particular, this proposal includes the generalized global symmetries defined in \cite{Gaiotto:2014kfa}. When saturation occurs for any $R$
\be \label{comp}
\mathcal{A}(R)= \mathcal{A}(R')'\;,
\ee
there are no generalized symmetries, and the theory is complete. Intuitively, it has enough charged operators to break non-local operators into local ones. For ball-like regions (\ref{comp}) is called Haag duality and it is expected to hold for quite general QFT's \cite{haag2012local}.

It turns out that given the previous inclusion of algebras (\ref{caus}), we have a dual inclusion of algebras as well \cite{Magan:2020ake,Casini:2020rgj}, associated with the complementary region
\be \label{caus1}
\mathcal{A}(R')\subset \mathcal{A}(R)'\;.
\ee
This just follows by taking the commutant of the previous one. Each of the inclusions (\ref{caus}) and (\ref{caus1}) is not saturated due to the existence of certain ``non-local'' operators that commute with local degrees of freedom in the complementary region, but cannot be generated locally by degrees of freedom in the region itself. These dual non-local operators exist based on complementary regions. They are the ``order'' and ``disorder'' parameters for the generalized symmetries. One of the advantages of this approach is that it naturally allows constructing entropic order parameters defined using relative entropy \cite{Magan:2020ake,Casini:2020rgj}. 

Given these observations, the objective of this article is to analyze the concrete and important example of generalized symmetries for non-Abelian gauge theories in the weak coupling limit following the approach of \cite{Casini:2020rgj}. In the present case, this means to find expectation values for smeared non-local operators and compute the associated entropic parameters, within a certain approximation. On the other hand, in displaying the behavior of these relative entropies in the weakly coupled regime we accomplish a necessary preliminary step in understanding the behavior of these parameters with the renormalization group. 

As it is well known, the group of generalized symmetries for non-Abelian theories is a finite and Abelian group. For example, for a pure non-Abelian gauge theory with gauge group $G$ in $d=4$ this group is $Z\times Z^* $, where $Z$ is the center of $G$, and $Z^*$ its dual group. In contrast, the Maxwell field has a continuous group $\mathbb{R}\times \mathbb{R}^*$ of non-local operators. The size of the symmetry group is sensed by the entropic order and disorder parameters, whose sum is constrained to be the logarithm of the number of non-local operators classes. Indeed, as shown in \cite{Magan:2020ake,Casini:2020rgj,Hollands:2020owv}, for pure gauge theories we have
\be \label{cer}
S_{\textrm{order}}+S_{\textrm{disorder}}=\log |Z|\;,
\ee
where $|Z|$ is the order of the center $Z$ of the gauge group. Such relation, to be reviewed below, is called entropic certainty relation. It gives a constraint on the statistics of dual generalized symmetries corresponding to complementary regions. Here we put it to test in weakly coupled gauge theories.

At a technical level, to compute bounds on these entropic order parameters, we need to find \emph{smeared} versions of the non-local operators. In the present case, this means we should find smeared versions of Wilson loops (WL) and 't Hooft loops (TL) in non-Abelian gauge theories. Line operators usually considered in the literature are too singular to provide useful bounds. Besides, they have information on two disparate physical scales given by the width and the radius of the loop.  
Constructing useful smeared loop operators for non-Abelian gauge theories turns out to be a complicated problem that has not yet been satisfactorily solved, see \cite{Narayanan_2006,Lohmayer:2011si} for a proposal. In this regard, the use of relative entropy will prove very useful in allowing us to treat the problem in an extended non-gauge invariant Hilbert space. There, the construction of smeared class operators turns out to be much simpler. We find that Wilson and 't Hooft loops are labeled by the weight and co-weight lattice of the gauge group, respectively.\footnote{Typically, Wilson line operators are labeled by representations of the gauge group, while 't Hooft line operators are labeled by representations of the GNO dual \cite{Kapustin_2006}. Equivalently, this means they are labeled by \emph{dominant} weights and co-weights. The smeared versions we propose in this article are labeled instead by the \emph{full} weight and co-weight lattices.  Appendix \ref{rept} contains a brief review of group representation theory for the convenience of the reader.}
We provide a generic expression of these operators valid at any coupling and get the expectation values in the weakly coupled regime. These operators (their smearing functions) are then optimized to provide the strongest bounds. 

The expectation values are enough to determine the leading behavior of the order parameters and verify the certainty relation at weak coupling. Among other things, we find that smeared WL, as opposed to the infinitely thin ones (line operators), satisfy a constant law with maximal expectation value in the weak coupling limit, rather than a perimeter law.  

This implies that the entropic order parameter saturates to its maximal topological value, and we find the rate of approach to such value. On the other hand, the smeared TL does go to zero with a perimeter law for smeared thin loops. This implies the associated entropic disorder parameter has an exponential behavior. We find the coefficient of the perimeter law within a factor two accuracy.  We will see that the rates of approach to saturation are consistent with the certainty relation~(\ref{cer}). The results generalize also to scenarios in which we include charged matter that breaks part of the generalized symmetry. 

Since the output of the analysis is some coupling-dependent functions valid at weak coupling, we end up briefly discussing how they run with the renormalization group (RG) flow, by using the known results from beta functions in pure gauge theories. This gives important information on what can or can not be expected about the general behavior of entropic order parameters with the RG flow.

\section{Haag duality, symmetries and entropic order parameters}

 \begin{figure}[t]
\includegraphics[width=.5\linewidth]{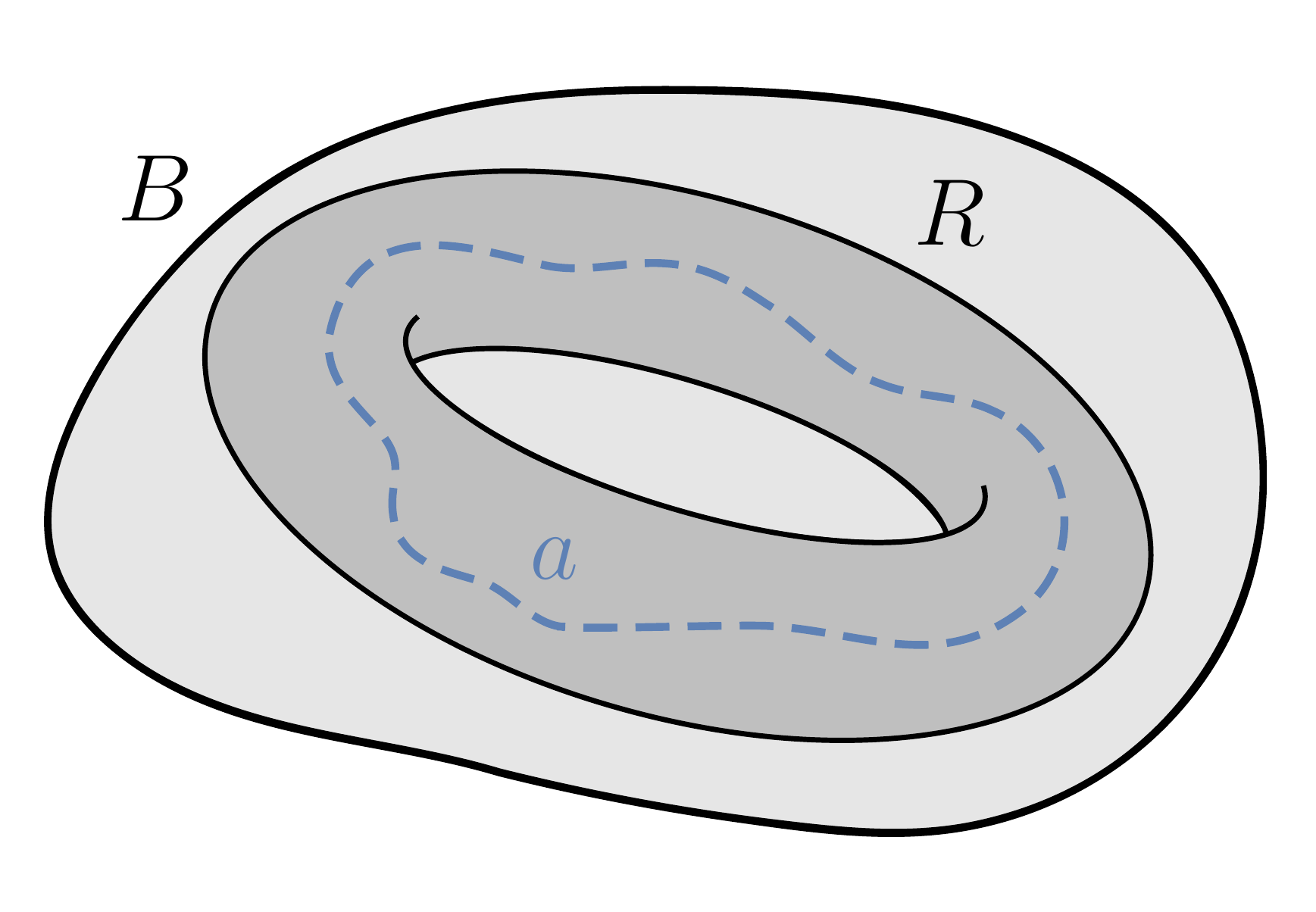}
\centering
\caption{We show an operator $a$ which is non-local in the topologically non-trivial region $R$. However, the same operator $a$ can be locally generated in the topologically trivial region $B$ containing $R$.}
\label{Non-local}
\end{figure}

When codifying the meaning of causality in the inclusion of algebras~(\ref{caus}), we have to specify what we mean with the algebra of a region, and this is of course crucial for the present purposes. Let us be more precise now and define ``the additive'' algebra of $R$  as
\be
{\cal A}_{\textrm{add}}(R)= \bigvee_{B \,\textrm{is a ball}, \,B\subseteq R} {\cal A}(B)\,.\label{ed} 
\ee
The operation ${\cal A}_1\vee {\cal A}_2$ between algebras means the algebra generated by taking arbitrary products of the two.
 Eq. (\ref{ed}) then provides a minimal algebra for a given region. It contains all operators which must form part of the algebra because they are locally generated by degrees of freedom in $R$. These are the operators which are physically accessible or measurable in $R$. This is for example the algebra of operators generated by the electric and magnetic fields in $R$ in the Maxwell theory.  The assignation of ${\cal A}_{\textrm{add}}(R)$ to any $R$  gives the minimal possible net of algebras for the QFT, called the additive net. With this definition, the previous causality constraint becomes more precise 
\be \label{caus2}
\mathcal{A}_{\textrm{add}}(R)\subset \mathcal{A}_{\textrm{add}}(R')'\;.
\ee
Consider a scenario in which this causality relation is not saturated for a region $R$ with non-trivial topology. We can equivalently say that duality is not satisfied for this region $R$. This results in two natural algebras associated with the same region $R$, namely
$\mathcal{A}_{\textrm{add}}(R)$ and $\mathcal{A}_{\textrm{add}}(R')'$. 
Since the second algebra is by construction the maximal possible algebra associated with $R$ and compatible with causality, we will rename it by
\be 
\mathcal{A}_{\textrm{max}}(R)\equiv\mathcal{A}_{\textrm{add}}(R')'\;.
\ee
Since $\mathcal{A}_{\textrm{add}}(R)$ is included in $\mathcal{A}_{\textrm{max}}(R)$, and moreover it is strictly smaller, it has to be the case that
\be 
\mathcal{A}_{\textrm{max}}(R)=\mathcal{A}_{\textrm{add}}(R)\vee \{a\}\;,
\ee
for some set $\{a\}$ of  operators  ``non-locally generated'' in $R$.   We will call the $a$ operators ``non-local operators'' for simplicity. However, it is important to keep in mind they while being non local in $R$, they actually can be locally generated in any topologically trivial region $B$ including $R$. We represent this in Fig. \ref{Non-local}.

 Von Neumann's double commutant theorem states that for any algebra the commutant of the commutant coincides with the algebra, ${\cal A}''={\cal A}$. Then, if ${\cal A}\subsetneq {\cal B}$ is an inclusion of different algebras, it must be the case that their commutants cannot be equal ${\cal B}'\subsetneq {\cal A}'$.   This leads to the conclusion that the strict inclusion of algebras for $R$ implies a strict inclusion of algebras for $R'$.
 This arises from taking the commutants of the two algebras associated with $R$. We have
\bea
\mathcal{A}_\textrm{add}(R') &=& \mathcal{A}_\textrm{add}(R')''=(\mathcal{A}_\textrm{add}(R')')'=(\mathcal{A}_\textrm{max}(R))'\,,\\
\mathcal{A}_\textrm{max}(R') &=& \mathcal{A}_\textrm{add}(R)'\;.
\eea
Hence, since the right hand sides are different, the left hand sides cannot coincide, $\mathcal{A}_\textrm{add}(R')\subsetneq\mathcal{A}_\textrm{max}(R')$. We conclude that the existence of non-local operators $a$ in $R$ forces the existence of non-local operators $b$ in $R'$, so that ${\cal A}_{\rm max}(R')={\cal A}\vee \{b\}$. These two set of non local operators $\{a\}$ and $\{b\}$ cannot commute with each other. In \cite{Magan:2020ake,Casini:2020rgj} this rigid structure was represented by means of the following complementarity diagram
\bea\label{cdiaor}
{\cal A}_{\textrm{add}}(R)\vee \{a\} & \overset{E}{\longrightarrow} &{\cal A}_{\textrm{add}}(R)\nonumber \\
\updownarrow\prime \!\! &  & \,\updownarrow\prime\\
{\cal A}_{\textrm{add}}(R')& \overset{E'}{\longleftarrow} & {\cal A}_{\textrm{add}}(R')\vee \{b\}\,.\nonumber 
\eea
In the upper part of this diagram, we have the two natural algebras associated with $R$. We can go from left to right employing a projection from the maximal to the additive subalgebra. These projections are called conditional expectations $E$. They are maps from the input to the target algebra leaving invariant the target algebra. We will explicitly define them below for the cases of interest in this paper. Going up and down in the diagram means taking commutants. In the lower part, we have the two algebras associated with the complementary region. Again, we can go from the bigger to the smaller algebras using another conditional expectation. Both conditional expectations are dual to each other, see \cite{Longo:1994xe,Magan:2020ake,Casini:2020rgj}.

Now, if region $R$ has non-trivial homotopy group $\pi_i$, the complementary region $R'$, i.e the set of points causally disconnected from $R$, has non-trivial $\pi_{d-2-i}$. 
The connection between this abstract framework and generalized symmetries arises as follows. The existence of the non-local operators ${a}$'s and $b$'s define classes/sectors in their respective regions. For example, the operators $a$ may be chosen to form irreducible classes $[a]$ in ${\cal A}_{\textrm{max}}(R)$ under multiplication by locally generated operators. More concretely, the subset $[a]$ of ${\cal A}_{\textrm{max}}(R)$ is  the set generated as
$\sum_\lambda O_1^\lambda \,a\, O_2^\lambda$, with $O_1^\lambda$ and $O_2^\lambda$ locally generated operators from ${\cal A}_{\textrm{add}}(R)$. These sectors satisfy certain fusion rules, and the same can be said for the complementary region $R'$. Further, the $a$ and $b$ operators do not commute with each other, and acting with the $a$ operators we can produce endomorphisms of the maximal algebra of the region $R'$ containing the $b$ operators, and vice-versa.  This means the non-local operators ${a}$'s and $b$'s associated with $R$ and $R'$ respectively, can be considered as ``topological operators'' effecting a generalized symmetry on their respective complementary regions $R'$ and $R$. The charged operators under these dual generalized symmetries are the topological operators $b$'s and $a$'s themselves.

The topological operators appearing in the definition of generalized global symmetries \cite{Gaiotto:2014kfa} are an example of this structure. In the applications of this paper, namely the case of generalized symmetries coming familiar from gauge theories, the dual fusion rules are associated with an Abelian group and its representations. The representations of an Abelian group form another Abelian group, known as the Pontryagin dual.

\subsection{Definitions of entropic order parameters  and bounds}
\label{def}

The causal incompleteness~(\ref{caus2}) associated with the existence of a generalized symmetry allows the introduction of entropic order and disorder parameters. In particular, given the previous complementarity diagram~(\ref{cdiaor}), the entropic order parameter is defined as
\be \label{or}
S_{\textrm{order}}\equiv  S_{{\cal A}_{\textrm{max}}(R)}(\omega,\omega\circ E)\;,
\ee
where $\omega$ is the vacuum state. 
 Analogously the entropic disorder parameter is defined by
\be \label{disor}
S_{\textrm{disorder}}\equiv  S_{{\cal A}_{\textrm{max}}(R')}(\omega,\omega\circ E')\;.
\ee
If there are subgroups of non-local operators various other analogous quantities can be considered as well.  
In these expressions, the entropies are relative entropies. They are functions of two states in the same algebra. The composition between a given state and a conditional expectation is defined in the obvious manner $\omega\circ E (\mathcal{O})\equiv\omega (E(\mathcal{O}))$, where $\omega({\cal O})$ is just the expectation value of the operator. The definition of relative entropy for matrix algebras, and more generally type I algebras, is 
\be 
S_{\mathcal{A}}(\rho,\sigma)=\textrm{Tr}_{\mathcal{A}}\rho\log \rho-\textrm{Tr}_{\mathcal{A}}\rho\log \sigma\;,
\ee
where $\textrm{Tr}_{\mathcal{A}}$ means the canonical trace associated with the algebra $\mathcal{A}$, and $\rho$, $\sigma$ are the two density matrices associated with the two states, see \cite{ohya2004quantum}.  The relative entropy exists as well for the algebras that appear in QFT (type III von Neumann algebras) \cite{ohya2004quantum}. Moreover, as shown below, when there is a finite number of non-local operators, both order and disorder parameters are always finite quantities in the continuum limit. 

Physically, these entropic order parameters give a measure of distinguishability between the two involved states. The state with the conditional expectation gives zero expectation value to all non-local operators. Therefore the entropic order parameters will increase when vacuum expectation values of non-local operators increase. They are, however, quantities depending on the geometry of $R$ and independent of the particular non-local operators we may choose to represent a certain non-local class. Heuristically, the relative entropies choose the best non-local operator with the largest expectation value.  

The challenge is of course to compute or approximate~(\ref{or}) and~(\ref{disor}) in actual QFT's. These are complicated quantities and we need to prescribe a procedure to organize the computation. There are two steps in this procedure. First, notice that we can obtain lower bounds to the previous relative entropies by using monotonicity of relative entropy
\bea
S_{{\cal A}_{\textrm{max}}(R)}(\omega,\omega\circ E)&\geq & S_{{\cal A}_{\textrm{lower}}(R)}(\omega,\omega\circ E)\,,\\
S_{{\cal A}_{\textrm{max}}(R')}(\omega,\omega\circ E')&\geq & S_{{\cal A}_{\textrm{lower}}(R')}(\omega,\omega\circ E')\;,
\eea
where ${\cal A}_{\textrm{lower}}(R)\subset {\cal A}_{\textrm{max}}(R)$ and ${\cal A}_{\textrm{lower}}(R')\subset {\cal A}_{\textrm{max}}(R')$ are conveniently chosen small algebras where we understand the expectation values of the operators. In order to complete the computation, it is very convenient to choose these new small algebras to be Abelian, but still including a complete set of non local operators. The non local operators are precisely the ones that are able to distinguish between the two states.  The fusion rules of the Abelian algebras can then be diagonalized, the output being a set of projectors $p_x$ running in a index $x$. We choose $x$ as a label because in the cases of interest in this paper they will arise from conventional discrete Fourier transforms. We can now compute the expectation values of the projectors $p_x$ in both states $\omega$ and $\omega\circ E$. Defining those as
\bea
p_x^{\omega} &\equiv &\omega (p_x)\,,\\
p_x^{E} &\equiv &\omega\circ E (p_x)\;,
\eea
the relative entropy becomes
\be \label{lowerbe}
S_{{\cal A}_{\textrm{max}}(R)}(\omega,\omega\circ E)\geq  S_{{\cal A}_{\textrm{lower}}(R)}(\omega,\omega\circ E)=\sum\limits_{x}p_x^{\omega}\log \frac{p_x^{\omega}}{p_x^{E}}\;,
\ee
and similarly for $R'$. Obviously, we need to choose the algebra ${\cal A}_{\textrm{lower}}(R)$ that provides the best lower bound, and that still allows to perform the computation.

Once we have computed both lower bounds, we would like to obtain upper bounds too. This is then given automatically by the certainty relation \cite{Magan:2020ake,Casini:2020rgj,Hollands:2020owv}. It relates the order and disorder parameters in a particular way
\be \label{uncert}
S_{{\cal A}_{\textrm{max}}(R)}(\omega,\omega\circ E)+S_{{\cal A}_{\textrm{max}}(R')}(\omega,\omega\circ E')=\log \lambda\;,
\ee
where, in a general scenario, $\lambda$ is the so-called index of the dual conditional expectations $E$ and $E'$, see \cite{Jones1983,KOSAKI1986123,longo1989,Longo:1994xe}. For the case of gauge theories, this index is simply the number of elements of the center of the unbroken part of the gauge group. Equivalently, it is the number of independent non-local operators in $R$ or $R'$. Given this relation, the previous lower bounds for the order and disorder parameters can be used to provide upper bounds for the disorder and order parameters respectively. This follows again from the monotonicity of relative entropy, together with the certainty relation. One obtains
\bea\label{boundsud}
S_{{\cal A}_{\textrm{lower}}(R)}(\omega,\omega\circ E)&\leq & S_{{\cal A}_{\textrm{max}}(R)}(\omega,\omega\circ E)\leq \log \lambda -S_{{\cal A}_{\textrm{lower}}(R')}(\omega,\omega\circ E')\,, \nonumber\\
S_{{\cal A}_{\textrm{lower}}(R')}(\omega,\omega\circ E')&\leq & S_{{\cal A}_{\textrm{max}}(R')}(\omega,\omega\circ E')\leq \log \lambda -S_{{\cal A}_{\textrm{lower}}(R)}(\omega,\omega\circ E)\;.
\eea
The task is then to choose the algebras ${\cal A}_{\textrm{lower}}(R)$ and ${\cal A}_{\textrm{lower}}(R')$ such that they make previous inequalities as tight as possible.

\section{A Maxwell field with ``$Z_{N}$-symmetry''}
\label{entropic}

As a warm-up exercise, and because the calculation is tightly related to the one in the next section for non-Abelian theories, we start by considering a subgroup $Z_N$ of generalized symmetries of the Maxwell field in $d=4$. This allows us to introduce some quantities and notation for the Maxwell field that will be of use later. The case of the subgroup of the integers  ${\mathbb Z}$ was considered in \cite{Casini:2020rgj}. The Maxwell field can be defined without the aid of the gauge potential, as the Gaussian theory of electric and magnetic fields satisfying
\be\label{ceb}
[E^i(\vec{x}),B^j(\vec{y})]=i \varepsilon^{ijk}\, \partial_k \delta^3(\vec{x}-\vec{y})\,.
\ee
Let us consider oriented electric and magnetic fluxes $\Phi_E$ and $\Phi_B$. These are defined on two-dimensional surfaces with boundaries $\Gamma_E$ and $\Gamma_B$. Because $\nabla E=\nabla B=0$, these fluxes are conserved. Indeed, the defining surface can be deformed, without modifying the operator, as long as we keep the boundary fixed. As a consequence, we can deform the surface of the flux to avoid any local operator lying on the original surface. Therefore,  fluxes based on a ring-like region necessarily commute with the locally generated operators associated with the complementary ring.   

By exponentiating and smearing the fluxes, we can write a bounded  electric flux operator, the TL $T^{\tilde{q}}=e^{i \tilde{q} \Phi_E}$, and a magnetic flux operator, the WL $W^{q}=e^{i q \Phi_B}$. These exist for any $\tilde{q},q\in \mathbb{R}$. When their respective boundaries are linked, the commutation relations between them follow from ~(\ref{ceb}) and are given by
\be
T^{\tilde{q}}W^{q}= e^{i\, q\, \tilde{q} } \, W^q T^{\tilde{q}}\,.\label{cr}
\ee 
This lack of commutativity for operators associated with causally disconnected rings implies these operators cannot be locally generated in those rings. For example, if $T^{\tilde{q}}$ were locally generated in $R$ (where its boundary $\Gamma_E$ lies) this would imply, by the arguments given above, it necessarily commutes with any $W^q$ based on the complementary ring $R'$ due to flux conservation. But this would contradict \eqref{cr}. Therefore, we can interpret the WL and TL operators as generators of a dual pair of generalized symmetries under which the TL and WL are charged, respectively.

We conclude that the Maxwell field violates Haag duality on rings, where, on top of the additive algebras  we have the non-locally generated WL and TL. In the generic nomenclature of the previous section
\bea
{\cal A}_{\textrm{max}} (R)\equiv ({\cal A}_{\textrm{add}}(R'))'={\cal A}_{\textrm{add}}(R)\vee \{W^q_{R} T^{\tilde{q}}_{R}\}_{q,\tilde{q} \in \mathbb{R}}\,,
\eea  
and analogously by interchanging $R\leftrightarrow R'$. Here we have denoted $W^q_{R}$ and $T^{\tilde{q}}_{R}$  for the bounded WL and TL based on $R$.

For the present article, the problem with this model is that there is an infinite number of non-local operators in each ring $R$. This implies that the certainty relation is less predictive since it says the sum of order and disorder parameters is infinite. We then seek for the smallest variation of this scenario in which we get a finite index, so that the certainty relation can be explored and verified.

\subsection{Complementarity diagrams}
\label{com}

For the Maxwell field (in $d=4$), ${\cal A}_{\textrm{add}}(R)$ is the additive algebra of the electric and magnetic fields inside the ring. The non-local operators are the WL and TL wrapping the ring. Therefore, the canonical complementarity diagram is
\bea
{\cal A}_{WT}(R) & \overset{E_{WT}}{\longrightarrow} & {\cal A}_{\rm add}(R)\nonumber \\
\updownarrow\prime\! &  & \:\updownarrow\prime\\
\mathcal{A}_{\rm add}(R') & \overset{E'_{WT}}{\longleftarrow} & {\cal A}_{WT}(R')\;.\nonumber 
\eea
All non-local operators with any charge are included in ${\cal A}_{WT}(R)$ and ${\cal A}_{WT}(R')$. As mentioned before, this diagram reminds us that the symmetry group of the non-local operators in a ring forms an infinite non-compact group ${\mathbb R}^2$ of electric and magnetic charges. The order parameters measuring the difference between  ${\cal A}_{\textrm{max}}(R)={\cal A}_{WT}(R)$ and ${\cal A}_{\textrm{add}}(R)$ are divergent.

However, order parameters for any subgroups of non-local operators can be analyzed. The case of a discrete group ${\mathbb Z}$ was discussed in \cite{Casini:2020rgj}. One starts by adding to ${\cal A}_{\textrm{add}}(R)$ only a closed group of WL corresponding to charges that are integer multiples of a given $q$. We call this algebra
\be
{\cal A}_{W_q}(R)\equiv {\cal A}_{\rm add}(R) \vee \{W_q\}\;.
\ee
A generic element in this algebra reads
\be
\sum_{m\in\mathbb{Z}} a_m (W_{q})^m\;,\qquad\qquad a_m\in {\cal A}_{\rm add}(R)\;.
\ee
There is an associated entropic parameter $S_{{\cal A}_{W_q}}(\omega|\omega\circ E_W)$ for this choice of algebra, where $E_W$ eliminates the non-additive WL, as defined by
\be \label{cond1}
E_W\left(\sum_{m\in\mathbb{Z}} a_m (W_{q})^m\right)=a_0\;.
\ee
The previous complementarity diagram gets modified in the following way
\bea\label{cus}
{\cal A}_{W_q}(R) & \overset{E_W}{\longrightarrow} & {\cal A}_{\rm add}(R)\nonumber \\
\updownarrow\prime\! &  & \:\updownarrow\prime\\
\mathcal{A}_{W,T_{\tilde{q}}}(R') & \overset{E_T}{\longleftarrow} & {\cal A}_{WT}(R')\;,\nonumber 
\eea
The algebra $\mathcal{A}_{W,T_{\tilde{q}}}(R')$ contains all non-local WL, and the non-local TL whose charges are integer multiples of 
$$\tilde{q}:=2\pi/q\;.$$ 
These are the only ones that commute with the WL appearing in ${\cal A}_{W_q}(R)$. The difference between ${\cal A}_{W_q}(R)$ and  ${\cal A}_{\rm add}(R)$ is a group $\mathbb{Z}$ of WL, while the difference between ${\cal A}_{WT}(R')$ and $\mathcal{A}_{W,T_{\tilde{q}}}(R')$ is its dual group $U(1)$ of TL. We denote by $E_T$ the dual conditional expectation. It eliminates the TL with non-integer multiple of the minimal magnetic charge $\tilde{q}$.  Because these dual groups contain an infinite number of elements, the index is still infinite, and we still lose some predictive potential of the certainty relation.\footnote{It turns out that $S_{{\cal A}_{W_q}(R)}(\omega|\omega\circ E_W)$, arising from a discrete group, is finite. The complementary $S_{{\cal A}_{WT}(R')}(\omega|\omega\circ E_T)$ diverges, as it should given the divergence on the index, see \cite{Casini:2020rgj}.}

In this article, we consider yet another variation on the theme but having a finite index. We will still add to ${\cal A}_{\textrm{add}}(R)$ the same closed group of WL corresponding to charges that are integer multiples of a given $q$. This means that the left part of the complementarity diagram~(\ref{cus}) remains unchanged. Now, instead of using a conditional expectation that kills all such non-local WL, we take a conditional expectation that kills all WL whose charge is not divisible by $Nq$. In other words, instead of the target algebra $ {\cal A}_{\rm add}(R)$, which would correspond to an infinite group with infinite index, we add to such algebra the WL with charges $\pm N q, \pm 2 N q,\cdots$. We call this algebra $ {\cal A}_{W_{Nq}}(R)$.   Remembering the generic expression~(\ref{cond1}) for an element $\mathcal{O}\in  {\cal A}_{W_{q}}(R)$, this new conditional expectation is defined as
\be 
E_{Z_N}\left(\sum_{m\in\mathbb{Z}} a_m (W_{q})^m\right)=\sum_{m\in\mathbb{Z}} a_{Nm} (W_{q})^{Nm}\;.
\ee
This conditional expectation therefore only kills a $Z_N$ subgroup of the initially infinite group. Its index is finite and given by $N$.

The commutant of the algebra $ {\cal A}_{W_{Nq}}(R)$ contains all WL, but only those TL whose charges are multiples of $\tilde{q}/N=\frac{2\pi}{Nq}$, in order to satisfy the Dirac quantization condition. Following analogous terminology, we call this algebra $ {\cal A}_{W,T_{\tilde{q}/N}}(R)$. The dual conditional expectation  kills all those TL whose charge is not divisible by $\tilde{q}$. Therefore, this conditional expectation keeps those TL with charges multiples of $\tilde{q}=2\pi/q$. If we parametrize a generic element of $ {\cal A}_{W,T_{\tilde{q}/N}}(R)$ as
\be 
\sum_{m\in\mathbb{Z}} a_m (T_{\tilde{q}/N})^m\,,\,\,\,\,\,\,\,\,\,\,\,\,\,\,\,\,\, a_m\in {\cal A}_W (R)\;,
\ee
then this dual conditional expectation acts as
\be 
E_{Z_N}'\left(\sum_{m\in\mathbb{Z}} a_m (T_{\tilde{q}/N})^m\right)=\sum_{m\in\mathbb{Z}} a_{Nm} (T_{\tilde{q}/N})^{Nm}=\sum_{m\in\mathbb{Z}} a_{Nm} (T_{\tilde{q}})^{m}\;.
\ee
The associated complementarity diagram is
\newpage 
\bea\label{cus1}
{\cal A}_{W_q}(R) & \overset{E_{Z_N}}{\longrightarrow} &  {\cal A}_{W_{Nq}}(R)\nonumber \\
\updownarrow\prime\! &  & \:\updownarrow\prime\\
\mathcal{A}_{W,T_{\tilde{q}}}(R') & \overset{E_{Z_N}'}{\longleftarrow} & {\cal A}_{W,T_{\tilde{q}/N}}(R')\;.\nonumber 
\eea
The index associated with both $E_{Z_N}$ and $E_{Z_N}'$ is $N$, and the certainty relation  becomes 
\be 
S_{{\cal A}_{W_q}(R)}(\omega,\omega\circ E_{Z_N})+S_{ {\cal A}_{W,T_{\tilde{q}/N}}(R')}(\omega,\omega\circ E'_{Z_N})=\log N\;.
\ee
One nice feature of this scenario is that a similar relation appears for pure non-Abelian gauge theories when the center $Z$ of the gauge group has order $\vert Z\vert=N$.

\subsection{Smeared loops and their algebra for abelian fields}

Before proceeding, we need to find actual WL and TL {\sl bounded operators}. The infinitely thin ones (line operators) usually considered in the literature are highly singular and are not bounded. On the other hand, the generic arguments described above suggest there should be a bounded (or smeared) version of them. In the Abelian case, this problem has a simple solution, as we now show. But this will become a more complicated problem for non-Abelian fields.

Let's start with the WL for the Maxwell field. We can define a smeared version of them by 
\be 
W=e^{i\Phi_B}=e^{i\int d^3 x \,A_i J^i}\,,\quad \partial_i J^i =0 \,.
\label{wilsonloop}
\ee
It is enough to smear in three dimensions at $x^0=0$ in the free case. Taking $J$ of compact support, its conservation implies that the loop is indeed gauge invariant
\be 
A_i \to A_i + \partial_i \Lambda  \quad \Rightarrow \quad W\to We^{i\int d^3 x \,(\partial_i \Lambda) J^i} = We^{-i\int d^3 x \,\Lambda (\partial_i  J^i)} = W\,,
\ee
where in the last equality we have imposed the smooth current distribution $J$ to have compact support on a certain ring $R$.

Current conservation implies that the flux over a two-dimensional surface $\Sigma$ that cuts the ring $R$ once is independent of the particular cross section $\Sigma$. Such flux defines the dimensionless ``charge'' of the WL
\be 
q=\int_\Sigma d\sigma \,n_i\, J^i\;.
\label{qwilson}
\ee
The vector $n_i$ is the unit normal to $\Sigma$. 
Being gauge invariant and unitary for any $q$, this is a bounded operator of the Maxwell theory. Furthermore, being constructed in the ring through the gauge potential, it is also potentially a non-locally generated operator in the ring.
In any case, as happens with any operator of the theory, we should be able to construct it employing gauge invariant local operators located in a sufficiently big ball. To see this is the case, one can prove by direct computation that
\be 
\Phi_B=\int d^3 x \,A_i J^i = \int d^3x\, j_{i} B^i\,, \label{1}
\ee
where $j_{i}$ now has support in a ball containing $R$, and it is defined by
\be
J_k=-\epsilon_{ijk} \partial_j j_i  \,.
\label{2}
\ee
There are many solutions for $j_i$ from this equation, but they all give place to the same operator $\Phi_B$. 

We can proceed analogously with the TL operator, which is the flux of the electric field. It can be written in a dual way to the WL as
\be 
T= e^{i\Phi_E} = e^{i\int d^3x\, \tilde{j}_{j} E^j} = e^{i \int d^3 x \,\tilde{A}_i \tilde{J}^i}\,,
\label{thooftloop}
\ee
where 
\be
E_k=\epsilon_{ijk} \partial_i \, \tilde{A}_j\,,\hspace{.6cm} \tilde{J}_k=-\epsilon_{ijk} \partial_j \tilde{j}_i\,,\hspace{.6cm} \partial_i \tilde{J}_i=0\,.
\ee

Again, if the current $\tilde{J}$ has support on a ring, the monopole charge of the smeared TL can be measured by integrating its flux over a two-dimensional cross section $\tilde{\Sigma}$ with normal vector $\tilde{n}_i$
\be 
\tilde{q}=\int_{\tilde{\Sigma}} d\sigma\, \tilde{n}_i\, \tilde{J}^i\,.
\label{gthooft}
\ee

\begin{figure}[t]
\includegraphics[width=.5\linewidth]{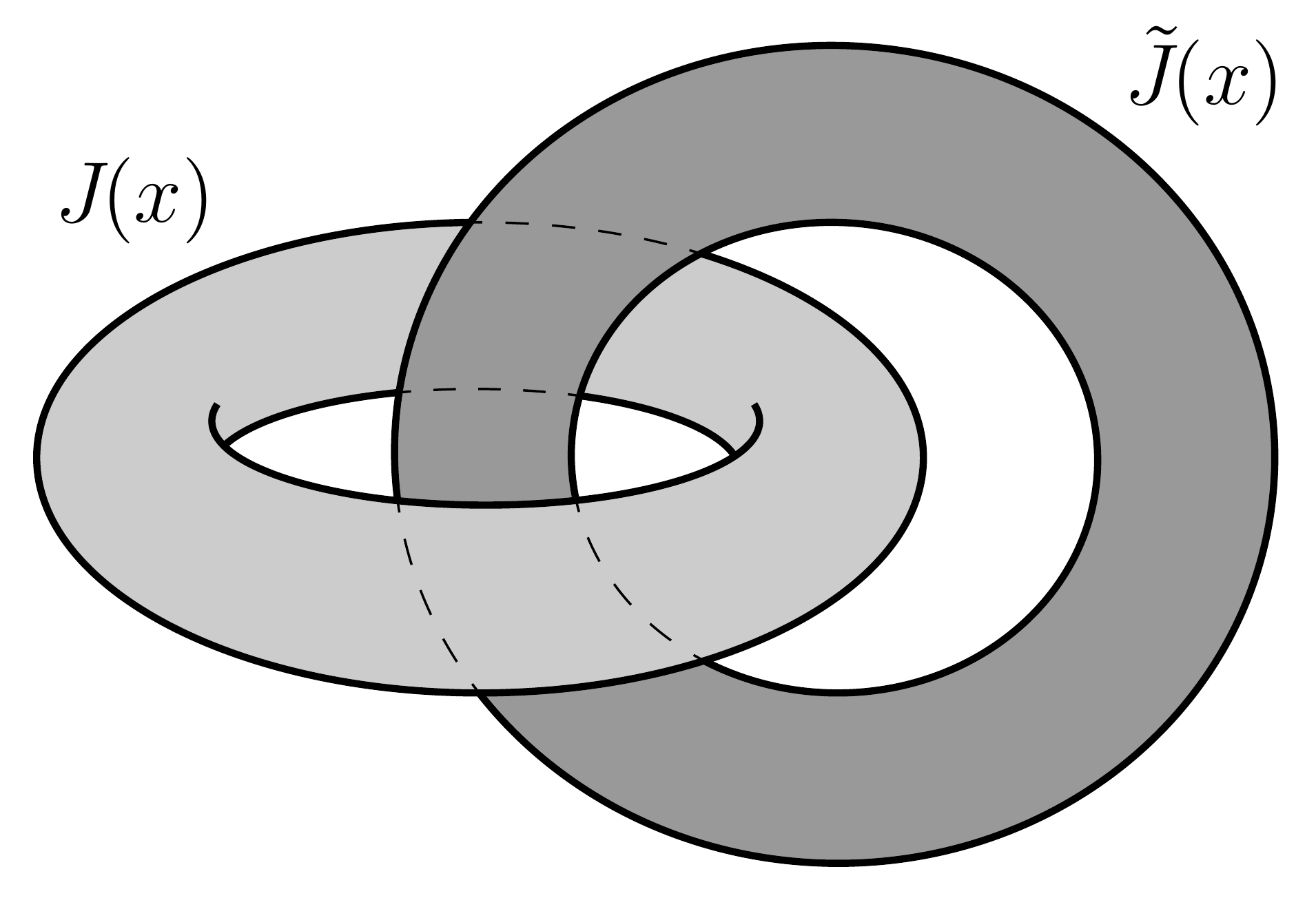}
\centering
\caption{The support of $J(x)$ and $\tilde{J}(x)$ for linked loops.}
\label{rings}
\end{figure}

When the supports of $W$ and $T$ are linked, as in figure \ref{rings}, we can find by direct computation that these smeared WL and TL operators satisfy~(\ref{cr}).

\begin{figure}[t]
\begin{center}  
\includegraphics[width=0.4\textwidth]{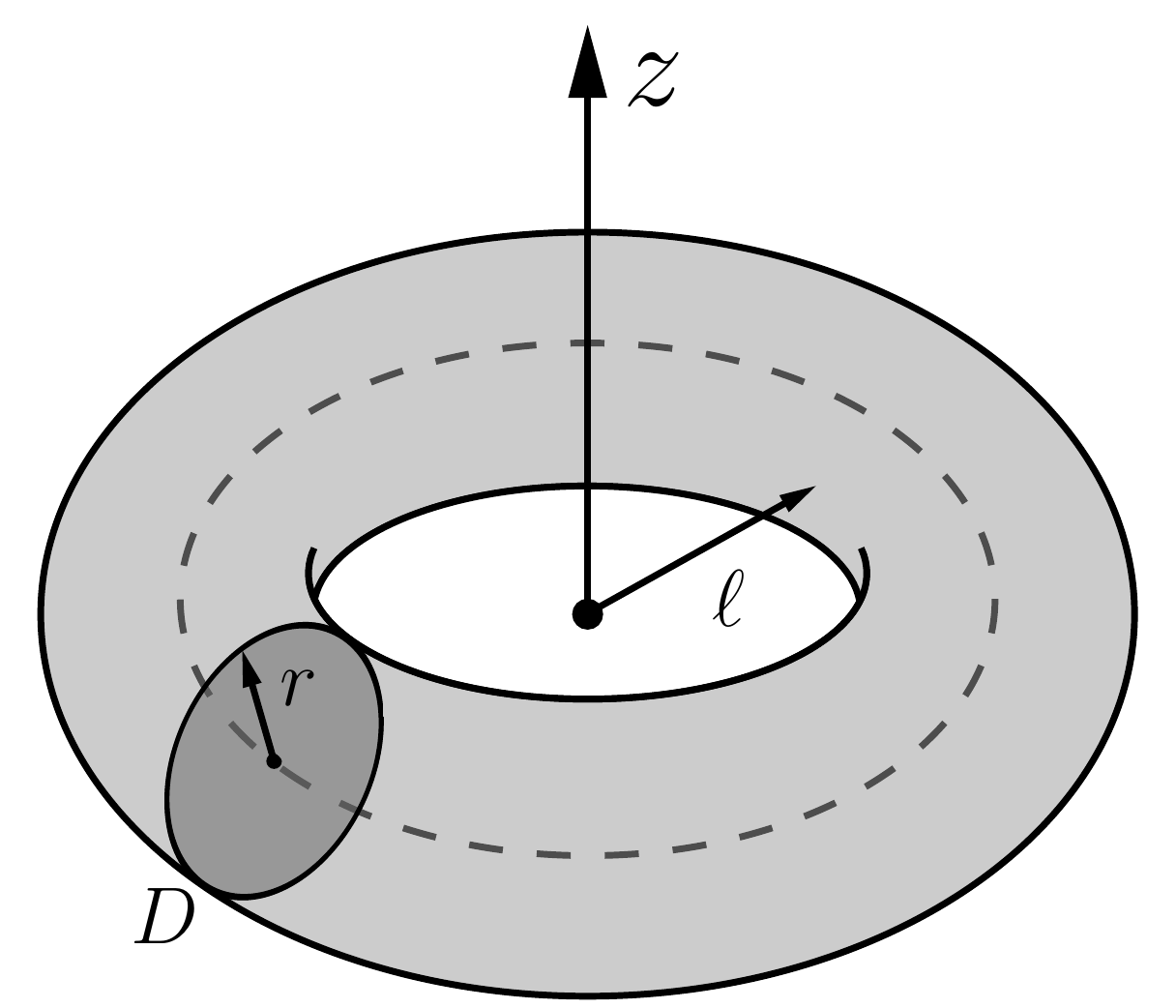}
\captionsetup{width=0.9\textwidth}
\caption{A ring formed by the revolution around the $z$ axis of a disk $D$ of radius $r$, such that the inner radius of the ring is $l$.}
\label{toro}
\end{center}  
\end{figure}

For concreteness, in what follows we will consider a ring with particular geometry formed by the revolution around the $z$ axis of a disk $D$ of radius $r$, such that the inner radius of the ring is $l$ (see figure \ref{toro}). A nice aspect of this symmetric ring is that, defining the cross ratio
\be  \label{crossr}
\eta=\frac{r^2}{(r+l)^2} \in (0,1)\;,
\ee
 the complementary region $R'$ is conformally equivalent to a ring with dual cross ratio \cite{Casini:2020rgj}
\be \label{crossrdual}
\eta '=1-\eta\;.
\ee

\subsection{Entropic order and disorder: exact  bounds and numerical evaluation}
\label{eno}

As explained above, to compute the entropic parameters, we first compute lower bounds to them by choosing appropriate subalgebras. Since the entropic parameters measure the difference between one state in which certain non-local operators have non-zero expectation values and another state in which these non-local operators have zero expectation values, a good choice of subalgebras is the ones generated by some non-local operators alone.

For the entropic order parameter, we can choose a given subalgebra of WL generated by the one of charge $q$. In the notation of the previous section, we are choosing
\be
{\cal A}_{\textrm{lower}}(R)=\{W_q\}\subset {\cal A}_{\textrm{max}}(R)\;.
\ee
The lower bound to the entropic order parameter is then
\be 
S_{W_q}(\omega,\omega\circ E_{Z_N})\;,
\ee
where we remind that the subindex means we are computing the entropy in the algebra $\{W_q\}$, generated by $W_q$. This is the Abelian algebra corresponding to the group ${\mathbb Z}$, namely
\be 
W_{q}^nW_{q}^m=W_{q}^{m+n}\equiv W^{m+n}\;.
\ee
It can be diagonalized via a traditional Fourier transform. Mathematically, we go from the group algebra (labeled by $m\in \mathbb{Z}$) to the character algebra (labeled by $x\in[-\pi,\pi]$), which is that of $U(1)$. Explicitly, the diagonalization of the algebra is accomplished by defining the projectors
\be 
p_x\equiv\,\frac{1}{2\pi}\,\sum_{m\in\mathbb{Z}}e^{imx}\, W^m \;.
\ee
It is easy to verify they are indeed projectors and they add up to one
\be 
\int_{-\pi}^{\pi} dx \,p_x=1\;.
\ee
We take a smeared WL of charge $q$, as defined in~(\ref{wilsonloop}). Actually, we will find more convenient to pull out a factor of $q$ in the WL definition and to constraint the flux of the current $J$ to be one. We also conveniently define
\be 
c^2=\frac 12 q^2 \langle\Phi_B^2\rangle\;,
\ee
 so that for a free field
\begin{equation}
W = e^{i q \Phi_B}
\quad \Rightarrow 
\quad 
\langle W^m \rangle = e^{-\frac 12 m^2 q^2 \langle\Phi_B^2\rangle}=e^{-m^2 c^2}\;.
\end{equation}
We can now compute the expectation values of the orthogonal projectors of this Abelian algebra. The expectation values in the vacuum were called previously $p_{x}^{\omega}$. They are given by
\be \label{w1Max}
p_{x}^{\omega}=\frac{1}{2\pi}\sum_{m\in\mathbb{Z}} e^{i m x} \langle W^m \rangle = \frac{ e^{-\frac{x^2}{4 c^2}}
   }{2\sqrt{\pi }c} \; \vartheta _3\left(-\frac{i x \pi }{2
   c^2},e^{-\frac{\pi
   ^2}{c^2}}\right)\;, 
\ee
where $\vartheta _3$ is the theta function. The expectation values in the vacuum composed with the conditional expectation were called $p_{x}^{E}$. They are given by
\be \label{w0Max}
p_{x}^{E}=\frac{1}{2\pi}\sum_{m\in\mathbb{Z}}e^{i N m x}\,\langle W^{N m} \rangle = \frac{e^{-\frac{x^2}{4 c^2}} }{2\sqrt{\pi }N c}\;
   \vartheta _3\left(-\frac{i x \pi }{2 N
   c^2},e^{-\frac{\pi ^2}{N^2
   c^2}}\right)\;.
\ee
The relative entropy is just given by
\begin{equation}\label{wSREL}
S_{W_q}(\omega,\omega\circ E_{Z_N})=\int_{-\pi}^\pi dx \,p_{x}^{\omega}\,\ln \left(\frac{p_{x}^{\omega}}{p_{x}^{E}}\right) \;,
\end{equation}
which we evaluate numerically below.

For the lower bound associated with the entropic disorder parameter, we can choose a given subalgebra of TL generated by the one of charge $\tilde{q}/N$. In the notation of the previous section, we are choosing
\be
{\cal A}_{\textrm{lower}}(R')=\{T_{\tilde{q}/N}\}\subset {\cal A}_{\textrm{max}}(R')\;.
\ee
The lower bound to the entropic disorder parameter is then
\be 
S_{T_{\tilde{q}/N}}(\omega,\omega\circ E'_{Z_N})\;,
\ee
where  we are computing the entropy in the algebra generated by $T\equiv T_{\tilde{q}/N}$, with $\tilde{q}=2\pi/q$. This is again an abelian algebra corresponding to the group ${\mathbb Z}$. It can also be diagonalized via a Fourier transform, exactly as above. The projectors are
\be 
\tilde p_x\equiv\,\frac{1}{2\pi}\,\sum_{m\in\mathbb{Z}}e^{imx}\, T^m \;.
\ee
Using the notation 
\be\tilde c^2=\frac 12 \left(\frac{2\pi}{Nq}\right)^2 \langle\Phi_E^2\rangle\,,\ee
 and the smeared TL operator defined in~(\ref{thooftloop}), we have
\begin{equation}
T = e^{i \frac{2\pi}{Nq} \Phi_E}
\quad \Rightarrow \quad 
\langle T^m \rangle = e^{-\frac 12 m^2 \left(\frac{2\pi}{Nq}\right)^2 \langle\Phi_E^2\rangle}=e^{-m^2 \tilde c^2}\;,
\end{equation}
Doing the Fourier transform in order to find the projectors we find 
\begin{equation}\label{t1Max}
\tilde p_x^{\omega}=\,\frac{1}{2\pi}\,\sum_{m\in\mathbb{Z}}e^{i m x}\langle T^m \rangle = \frac{ e^{-\frac{x^2}{4 \tilde{c}^2}}
   }{2\sqrt{\pi }\tilde{c}}\; \vartheta _3\left(-\frac{i x \pi }{2
   \tilde{c}^2},e^{-\frac{\pi
   ^2}{\tilde{c}^2}}\right)\;, 
\end{equation}
in the vacuum state, and
\begin{equation}\label{t0Max}
\tilde p_x^{E}=\,\frac{1}{2\pi}\,\sum_{m\in\mathbb{Z}}e^{i N m x}\langle T^{Nm} \rangle= \frac{e^{-\frac{x^2}{4 \tilde{c}^2}}}{2\sqrt{\pi }N \tilde{c}}\;
   \vartheta _3\left(-\frac{i x \pi }{2 N
   \tilde{c}^2},e^{-\frac{\pi ^2}{N^2
   \tilde{c}^2}}\right)\;,
\end{equation}
in the vacuum state composed with the conditional expectation. Eqs. (\ref{t1Max}) and (\ref{t0Max}) are equal to (\ref{w1Max}) and (\ref{w0Max}) except for the replacement $c\rightarrow \tilde{c}$. We finally obtain
\begin{equation}
S_{T_{\tilde{q}/N}}(\omega,\omega\circ E'_{Z_N})=\int_{-\pi}^\pi dx \,\tilde p_x^{\omega}\,\ln \left(\frac{\tilde p_x^{\omega}}{\tilde p_x^{E}}\right)\;.
\end{equation}

Having computed the lower bounds we can now use relations~(\ref{boundsud}) to find the upper bounds. However, up until this point, we have computed lower bounds in terms of parameters $c,\tilde c$ which are themselves functions of the fluxes $\langle\Phi_B^2\rangle$ and $\langle\Phi_E^2\rangle$ respectively. In our problem, these fluxes are ultimately dependent on the specific $J$ and $\tilde J$ smearing functions given for the corresponding ring, see \eqref{1} and \eqref{thooftloop}. Therefore, providing optimal bounds for the theory under study corresponds to an optimization problem for the sources $J,\tilde J$.

\begin{figure}[t]
\includegraphics[width=.6\linewidth]{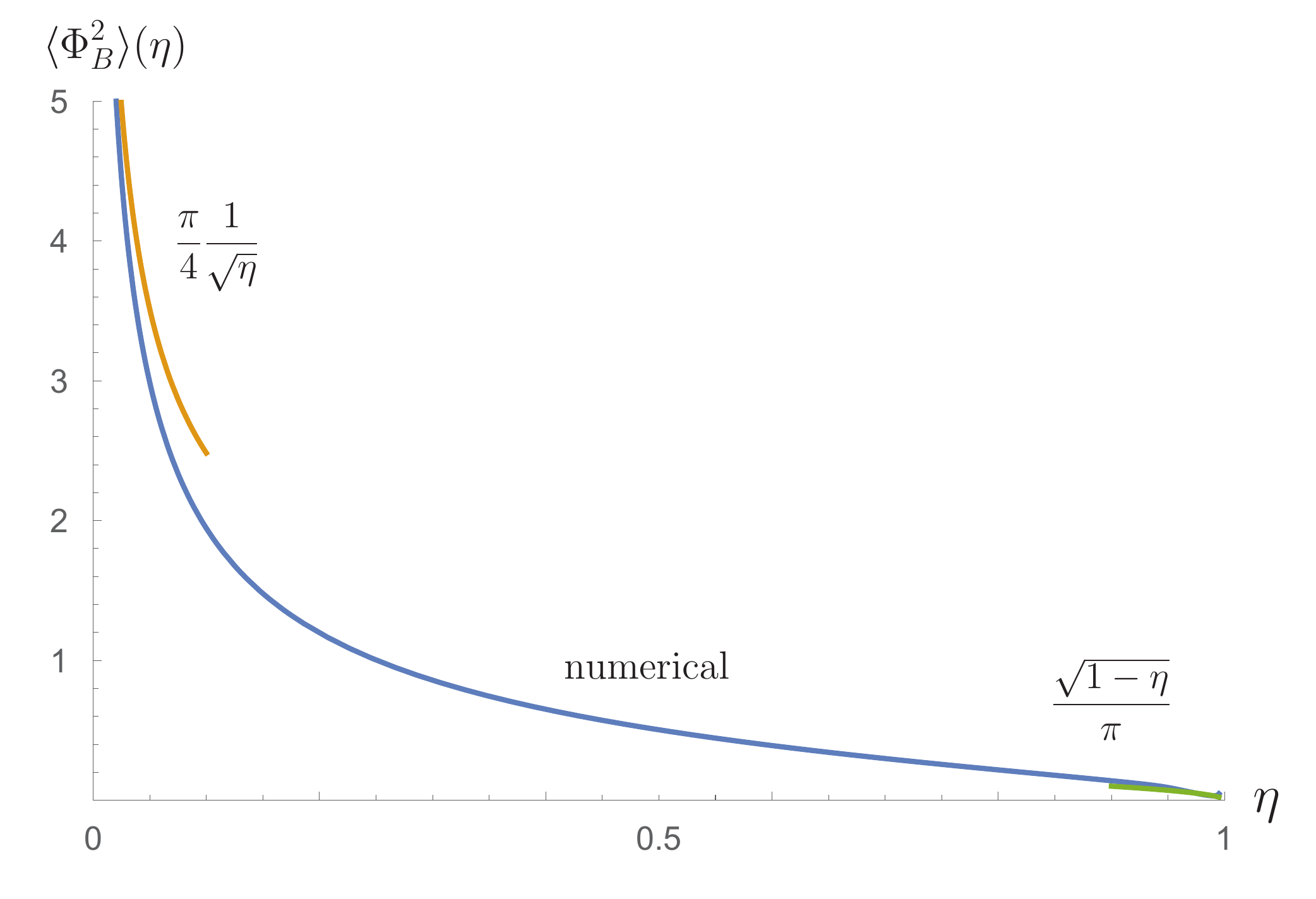}
\centering
\caption{Optimal flux squared as a function of the ring cross ratio.}
\label{FigFlux}
\end{figure}

As discussed in \cite{Casini:2020rgj}, optimal bounds are provided by $J,\tilde J$'s that minimize both flux squared and maximize the expectation values of non local operators. We take the rotationally symmetric ring described previously. We begin by finding $J$ such that it minimizes $\langle\Phi_B^2\rangle$ by choosing a smearing at a single instant $t=0$ and aligned with the angular direction $J= J_{\varphi}(r,z)\hat\varphi $.  The optimization problem is thus reduced to find a single function $ J_{\varphi}$ for a loop with unit charge \eqref{qwilson}, so that
\begin{equation}
\int_D du \; J_{\varphi}(u)\equiv J_{\varphi}\cdot1=1\;,
\end{equation}
where $u$ collectively denotes the coordinates $r,z$ on the disk $D$. This function  has to minimize the flux self-correlation
\begin{equation}
\langle\Phi_B^2\rangle=\int_D du\int_D du'  J_{\varphi}(u) K(u,u') J_{\varphi}(u')\equiv J_{\varphi}\cdot K\cdot  J_{\varphi}
\end{equation}
where $K(u,u')$ is the field's free correlator in cylindrical coordinates upon integration on $z$, see \cite{Casini:2020rgj}. 
This problem can be formally solved by
\begin{equation}
 J_{\varphi}=\frac{K^{-1}\cdot1}{1\cdot K^{-1}\cdot 1} \qquad\Rightarrow\qquad \langle\Phi_B^2\rangle=\frac{1}{1\cdot K^{-1}\cdot 1}\,.
\end{equation}
The optimal flux can be obtained by discretizing the kernel $K$ and taking a continuum limit.  A plot for the optimized flux is presented in Fig. \ref{FigFlux}.
Notice that by conformal invariance, $\langle\Phi_B^2\rangle$ can only be a function of the ring cross-ratio
\begin{equation}
\eta=\frac{r^2}{(r+l)^2}\,.
\end{equation}
Beyond the numerical solution, as shown in \cite{Casini:2020rgj} the limits $\eta\ll1$ and $1-\eta\ll 1$ admit analytic solutions
\begin{equation}
\langle\Phi_B^2\rangle(\eta\ll1)\sim\frac{\pi}{4}\frac{l}{r}=\frac{\pi}{4}\frac{1}{\sqrt{\eta}}\,,\qquad\qquad\langle\Phi_B^2\rangle(1-\eta\ll1)\sim\frac{1}{\pi}\sqrt{\frac{2l}{r}}=\frac{\sqrt{1-\eta}}{\pi}\,.\label{masc}
\end{equation}

By electromagnetic duality, the optimization problem for $\langle\Phi_E^2\rangle$ follows in an entirely analogous fashion.
We just need to notice that the lower bound to the entropic disorder, used to provide an upper bound to the entropic order, is computed in the complementary ring $R'$ with cross-ratio $\eta'$. Since complementary rings have a mirrored cross-ratio $\eta'=1-\eta$, we get the following simple relation for the optimal fluxes
\begin{equation}
\langle\Phi_E^2\rangle(\eta')=\langle\Phi_B^2\rangle(1-\eta)\,.
\end{equation}

In the free Maxwell theory, the normalized fluxes satisfy $[\Phi_B,\Phi_E]=1$ when they are linked to each other. Such a relation imposes an uncertainty relation bound for the fluxes 
\be \langle\Phi_B^2\rangle \langle\Phi_E^2\rangle\geq 1/4\,.\label{unc}
\ee
 The saturation corresponds to a pure state for the non-commuting Gaussian electric and magnetic modes. 
One can check that the numerical optimal solution lies very close to this theoretical limit, giving $\langle\Phi_B^2\rangle \langle\Phi_E^2\rangle\sim 0.36$ at its worst. The analytical approximations at the limits of very thin $\eta\rightarrow 0$ or wide $\eta\rightarrow 1$ loops saturate this bound.   Tighter bounds are then obtained in these limits. This means that no significant improvement is possible through flux optimization. To further improve the bounds one has to enlarge the algebras including additive operators in the rings. 

\begin{figure}[t]
\includegraphics[width=.6\linewidth]{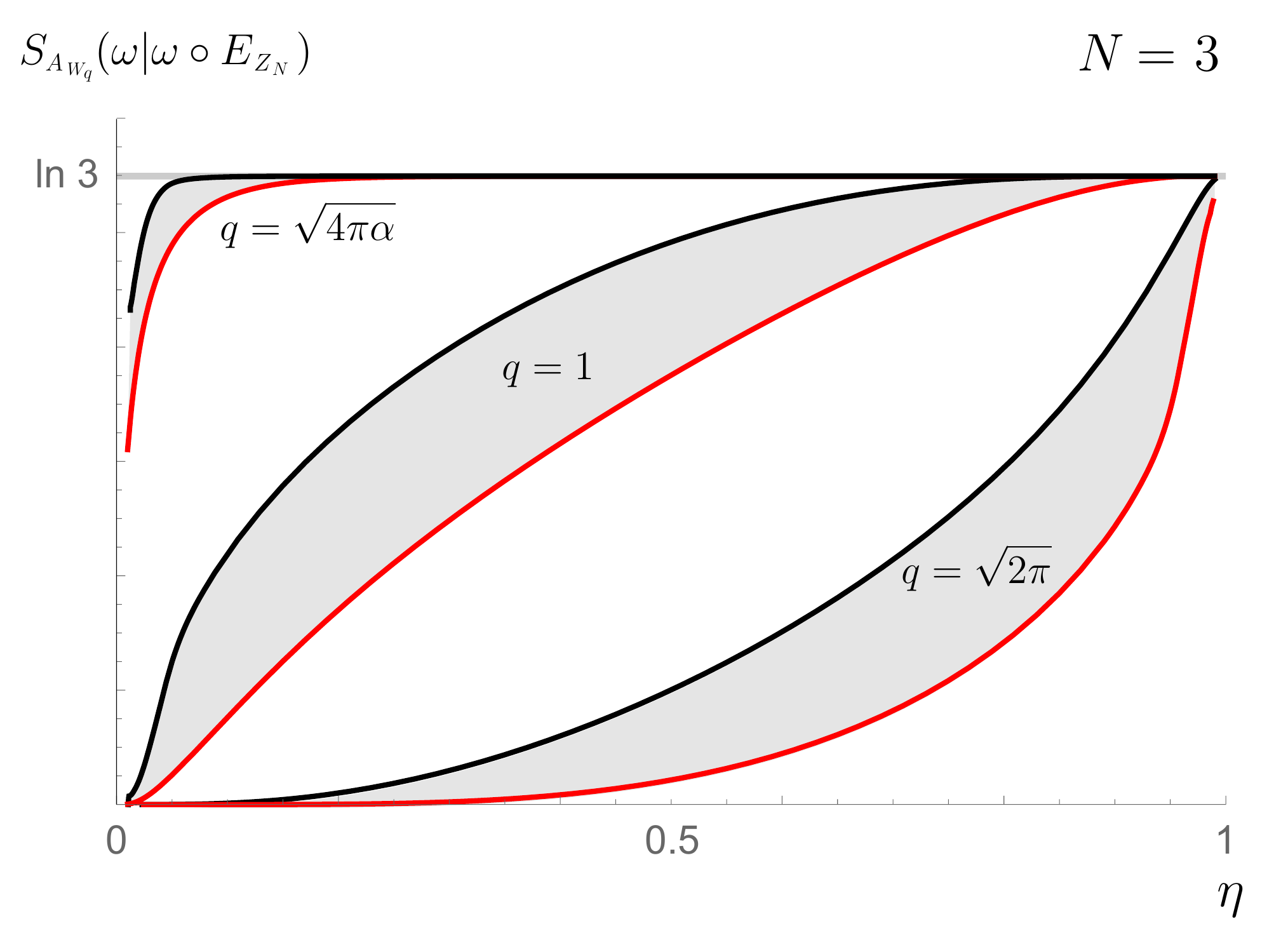}
\centering
\caption{Numerical upper and lower bounds for $S_{A_{W_q}}(\omega|\omega\circ E_{Z_N})$ with $N=3$ for 3 different charges: electron charge $q=\sqrt{4 \pi \alpha}$, $1$, and the self dual point $q=\sqrt{2 \pi}$.}
\label{MaxBounds}
\end{figure}

By putting all  this together we can provide upper and lower bounds for $S_{A_{W_q}}(\omega|\omega\circ E_{Z_N})$ in these Maxwell $Z_N$ models. Eq.\eqref{boundsud} takes the following form
\begin{equation}
S_{W_q}(\omega|\omega\circ E_{Z_N})\leq S_{A_{W_q}}(\omega|\omega\circ E_{Z_N})\leq \ln(N)- S_{T_{\tilde{q}/N}}(\omega|\omega\circ E_{Z_N})\,.
\end{equation}
We show an explicit computation for $N=3$ and some particular choices of $q$ in Fig. \ref{MaxBounds}. The results are compatible with the certainty relation.

\subsection{Entropic order parameters at weak coupling}
\label{analytic}

It is possible, and useful for the later analysis of non-Abelian gauge theories, to find regimes in which the previous elliptic formulas can be simplified and the entropic order parameters can be computed analytically. It will come as no surprise that this can be done in the weak coupling regime.

In this Gaussian theory, by ``weak coupling'' we mean small $q$. The expectation values for the projectors associated with the WL simplify to, $c^2\ll1$,
\be
p_x^{\omega}
\sim \sum_{n\in\mathbb{Z}}\frac{e^{-\frac{(x-2\pi n)^2}{4 c^2}}}{2 \sqrt{\pi } c}
\sim
\frac{e^{-\frac{x^2}{4 c^2}}}{2 \sqrt{\pi } c}\;,\label{polo}
   \ee
\be
p_x^{E}
\sim \sum_{n\in\mathbb{Z}}\frac{e^{-\frac{\left(x-\frac{2\pi n}{N}\right)^2}{4 c^2}}}{2 \sqrt{\pi } N c}
\sim \sum_{n=-\lfloor N/2 \rfloor}^{\lfloor N/2 \rfloor}\frac{e^{-\frac{\left(x-\frac{2\pi n}{N}\right)^2}{4 c^2}}}{2 \sqrt{\pi } N c}	
   \;.\label{pio}
\ee
where $\lfloor a \rfloor$ takes the integer part of $a>0$ and in the rhs we dropped all contributions lying outside of the $x\in[-\pi,\pi]$ domain in this limit. The entropic order parameter then takes the following form
\be
S_{W_q}(\omega|\omega\circ E_{Z_N})   \simeq
\ln (N)-\frac{4 c N }{\pi ^{3/2}}e^{-\frac{\pi ^2}{4 c^2 N^2}}=
\ln (N)-\frac{2 q N }{\pi ^{3/2}}\sqrt{2 \langle\Phi_B^2\rangle}e^{-\frac{\pi ^2}{2 N^2 q^2 \langle\Phi_B^2\rangle}}\;. \label{plo}
\ee
For the entropic disorder parameter, small $q$ amounts to $\tilde c^2\gg1$ and the expectation values for the projectors become
\be
\tilde p_x^{\omega}\sim \frac{1}{2\pi}+\frac{1}{\pi} e^{-\tilde{c}^2} \cos (x)\;,
\ee
\be 
\tilde p_x^{E}\sim \frac{1}{2\pi}+\frac{1}{\pi} e^{-N^2 \tilde{c}^2} \cos (x N)\;.
\ee
The entropic disorder parameter simplifies to
\be 
S_{T_{\tilde{q}/N}}(\omega|\omega\circ E_{Z_N})\simeq e^{-2 \tilde{c}^2}=e^{-\left(\frac{2 \pi }{N q}\right)^2 \langle\Phi_E^2\rangle}\;.\label{STMax222}
\ee

The certainty relation requires that (\ref{STMax222}) and (\ref{plo}) add up to less than $\ln N$. Considering the exponentials in the corrections this is true as long as  $\langle\Phi_B^2\rangle \langle\Phi_E^2\rangle\ge 1/8$, which is certainly obeyed by the quantum uncertainty relation (\ref{unc}), missing it by a factor $2$. In the limit of thin loops where (\ref{unc}) saturates and the approximations (\ref{masc}) are valid, the upper and lower bounds give the exact exponents in the corrections within a factor $2$.    

Before moving on, we comment on the underlying structure in the computation of the expectation values of the projectors, and consequently on the relative entropy itself, in the small coupling limit. This structure is not different for non-Abelian groups.

For the TL parameter, in the $q\ll 1$ limit, and to leading exponential order, the only relevant contribution to the sums in \eqref{t1Max} and \eqref{t0Max} are its first terms. Moreover, one can check that the first term in \eqref{t0Max} will always come exponentially suppressed by definition with respect to the one in \eqref{t1Max}, so that the first term in \eqref{t1Max} alone is enough for our purposes. 
That is, one only needs the expectation value of the TL with the smallest charge to get the leading order contribution to \eqref{STMax222}.

\begin{figure}[t]
\includegraphics[width=.6\linewidth]{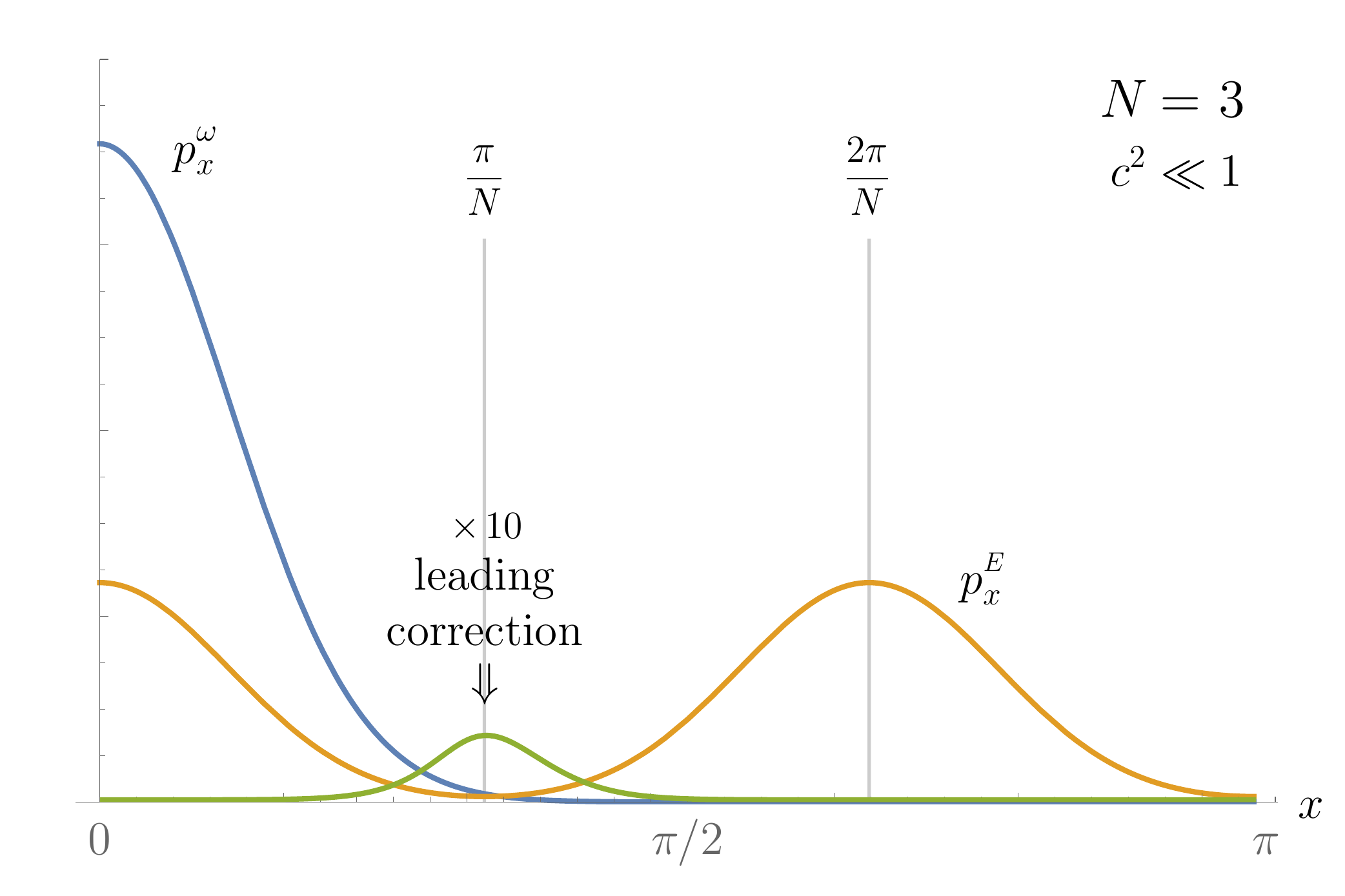}
\centering
\caption{$p_x^{E}$ has a structure of several peaks in $x\in[-\pi,\pi]$ obtained by multiplying $p_x^{\omega}$ by $1/N$ and adding displaced copies with periodicity $2\pi/N$. The comparison of the two peaks of these distributions at $x=0$ give the main $\log N$ contribution. The leading correction comes from the comparison between the $x=0$ peak of $p_x^{\omega}$ and the first displaced peak in $p_x^{E}$ at $x=2\pi/N$. The main contribution to the relative entropy correction (whose density is plotted in green) is located half way between these peaks. The $N=3$ case is shown in the figure. The functions are even in $x$ so we only plot for $x\in[0,\pi]$. }
\label{Fig:Gauss}
\end{figure}

For the WL parameter, taking the $q\ll 1$ limit directly in~(\ref{w1Max}), the sum can be approximated via an integral which leads to the rhs of (\ref{polo}). In this limit, the $p_x^\omega$ becomes a Gaussian.
As $c^2\to 0$, the expectation values $\langle W^m \rangle$ become indistinguishable from $W^{0}=1$, up until some large $m\gg1$. This implies the contributions for $x\sim 0$ dominate the Fourier transform. To obtain the leading exponential corrections\footnote{ In the approximation of the sum by an integral, i.e. using the Euler-MacLaurin formula, one misses these corrections completely and all the corrections to the Gaussian vanish order by order. The reason for this is that the Euler-MacLaurin formula is unable to capture exponentially small non-perturbative corrections.}  in (\ref{polo}) one notices that the character $e^{i m x}$, and, as a consequence the projector expectation value, must be periodic under $x\sim x+2\pi$. 
For $p_x^{\omega}$, these secondary Gaussians are never in $x\in[-\pi,\pi]$ and one may in fact disregard these contributions to leading order in this limit, as we did in \eqref{polo}. However, the analogous corrections in $p_x^{E}$ play a central role in the calculation of~(\ref{plo}). 
The $x\sim 0$ peak and periodicity structure also holds for $p_x^{E}$, but we have essentially replaced a $2\pi$ by a $2\pi/N$ periodicity, see (\ref{pio}).
Thus, $p_x^{E}$ is also a sum of Gaussians but with shorter periodicity, many of them ($2\lfloor N/2 \rfloor+1$ to be precise) now lying in $x\in[-\pi,\pi]$. Notice these Gaussians have a $1/N$ prefactor to keep $p_x^{E}$ correctly normalized. Of all these, only the ones at $\pm 2\pi/N$ are relevant to compute the leading correction in~(\ref{plo}). 
The quotient $N p_x^E/p_x^\omega$ in \eqref{wSREL} can be seen to be a sum of exponentials linear in $x$, which for $c\ll1$ dominate the sum only one at a time. Thus, its logarithm $\ln(N p_x^E/p_x^\omega)$ can be fairly approximated by an ensemble of linear functions, the logarithm of the dominating exponential, valid in the domain in which each exponential dominates the sum. The convolution of these linear functions with $p_x^\omega$ contains several contributions. However, the leading ones can be easily tracked down to the nearest Gaussians in $p_x^E$ at $\pm 2\pi/N$. The exponent in the leading correction to \eqref{wSREL} is given by the exponent in $p_x^{\omega}$ evaluated at half distance to the nearest peak in $p_x^{E}$, i.e. at $x=\frac \pi N$. We present a summary of this discussion in Fig. \ref{Fig:Gauss}.  

\section{Non Abelian gauge theories at weak coupling}
\label{sun}

Having studied a controlled example, we now move towards the case of interest in the paper, namely that of non-Abelian gauge theories.  We will first focus on pure gauge fields, with the usual Yang-Mills action
\be
{\cal L}=-\frac{1}{4} F_{\mu\nu}F^{\mu\nu}\;;\qquad\qquad F_{\mu\nu}^a=\partial_\mu A^a_\nu-\partial_\nu A^a_\mu + g f^{abc} A_\mu^b A_\nu^c \,,
\ee
where $F_{\mu\nu}^a$ is the non-Abelian generalization of the field strength for the Maxwell field.

The structure of algebras, and non-local operators, which manifest the generalized symmetries associated with these QFT's was analyzed in detail in Ref \cite{Casini:2020rgj}. As explained above, finding the potential generalized symmetries means finding those non-local operators in the theory which cannot be constructed by doing only local operations in certain topologically non-trivial regions. These operators then violate Haag duality of the additive net on those regions. For gauge theories, non-local operators violate duality in ring-like regions with non-trivial homotopy group $\pi_{1}$, and on the complementary regions with non-trivial $\pi_{d-3}$.

Naively, obvious candidates for the non-local operators violating Haag duality, i.e candidates for the generators of the generalized symmetries, are the Wilson and 't Hooft loops. But this has to be qualified. The reason is that one can prove \cite{Casini:2020rgj}, that for any gauge group $G$, whether continuous or discrete, the Wilson loop in the representation coming from the fusion  $rr^*$, where $r$ is any representation and $r^*$ is the conjugate representation, can be locally generated in a ring. It is thus not a proper non-local operator, in the sense that it does not violate Haag duality on the ring. The same can be said for any Wilson loop associated with any irreducible representation appearing in the decomposition of arbitrary powers of $rr^*$. In other words, to find the set of truly non-local Wilson loops, we have to quotient the full set of Wilson loops (one per irreducible representation) by the ones appearing in arbitrary products of $rr^*$. This quotient turns out to be isomorphic to the group of representations of the center of the gauge group $Z^*$, i.e the Pointyagrin dual of the center $Z$ of $G$. This is isomorphic to the center $Z$ itself. If the center has $|Z|$ elements, we then have $|Z|-1$ non-trivial non-local classes of Wilson loops, plus the identity. It is not a  coincidence that this is the same as the number of 't Hooft loops, as originally defined in Ref \cite{tHooft:1977nqb}, and which are labeled by elements of $Z$.

Therefore, as in the Maxwell case, we have two dual generalized symmetries, one generated by the non-local Wilson loops and the other by the 't Hooft loops. But in this case, the number of generators is finite and given by $Z$.\footnote{These generalized symmetries associated with the violation of Haag duality are the same as the generalized global symmetries discussed in \cite{Gaiotto:2014kfa}. The present approach allows us to find them algebraically, irrespective of the specific gauge invariant Lagrangian. This simplifies the consideration of subtle examples, such as discrete gauge groups.}

The violation of Haag duality and the existence of generalized symmetries could have been anticipated by the non-trivial commutation relation between Wilson and 't Hooft loops. For simply laced spatially separated rings
\be
T_z \, W_{z^*}= \chi_{z^*}(z) \, W_{z^*} \, T_z\,,\label{ccrr1}
\ee
where $z\in Z$, $z^*\in Z^*$, and $\chi_{z^*}$ is the character function. We remark again that the fact such dual sets of operators do not commute is not a violation of causality in QFT, but a demonstration that such operators cannot be locally generated in their respective linked rings.

\subsection{Complementarity diagrams}
\label{com1}

As described above, a violation of Haag duality implies the existence of two different algebras for the same region. We will focus on $d=4$, where the TL are also one dimensional loops. Therefore, on top of the additive algebras, for a ring we have the non-locally generated WL and TL, so that
\bea
{\cal A}_{\textrm{max}} (R)\equiv ({\cal A}_{\textrm{add}}(R'))'={\cal A}_{\textrm{add}}(R)\vee \{W^{z^*}_{R} T^z_{R}\}\,,
\eea  
and analogously by interchanging $R\leftrightarrow R'$. Therefore, for pure Yang-Mills the canonical complementarity diagram is
\bea
{\cal A}_{WT}(R) & \overset{E_{WT}}{\longrightarrow} & {\cal A}_{\rm add}(R)\nonumber \\
\updownarrow\prime\! &  & \:\updownarrow\prime\\
\mathcal{A}_{\rm add}(R') & \overset{E'_{WT}}{\longleftarrow} & {\cal A}_{WT}(R')\;.\nonumber 
\eea
All non-local operators are included in both ${\cal A}_{WT}(R)$ and ${\cal A}_{WT}(R')$. The symmetry group of the non-local operators in a ring is the discrete group $Z \times Z^*$ of electric and magnetic charges. The index of the inclusion is then $|Z|^2$. Since this group is finite and discrete, both entropic order parameters are finite. In fact, given the certainty relation, their sum equals $2\log |Z|$.

This entropic parameters in this complementarity diagram both depend on the WL and TL. In order to evaluate an order parameter in which we can measure the statistics of WL and that of TL separately it is natural to consider the analogue of the second complementarity diagram for the Maxwell field eq \eqref{cus}. In the present  case it reads
\bea
{\cal A}_{W}(R) & \overset{E_W}{\longrightarrow} & {\cal A}_{\rm add}(R)\nonumber \\
\updownarrow\prime\! &  & \:\updownarrow\prime\\
\mathcal{A}_{W}(R') & \overset{E_T}{\longleftarrow} & {\cal A}_{WT}(R')\;.\nonumber 
\eea
In other words, we add to the additive algebra of $R$ just the WL. The conditional expectation $E_W$ kills the WL to arrive at the additive algebra, and the dual conditional expectation $E_T$ just kills the TL in the complementary region $R'$. Several variations can be discussed, but we will focus on this complementarity diagram in what follows.

In this scenario, these conditional expectations can be defined by averaging over the dual generalized symmetry, namely
\bea
E_W(x_R)&=& \sum_{z\in Z} T^z \,x\,  (T^z)^{-1}\nonumber\\
E_T(x_{R'})&=& \sum_{z^*\in Z^*} W^{z^*} \,x\,  (W^{z^*} )^{-1}\;,\label{condW}
\eea
where $x_R\in {\cal A}_{W}(R)$ and $x_{R'}\in {\cal A}_{WT}(R')$. The index associated to both $E_{W}$ and $E_{T}'$ is $|Z|$ in this case, which is just the number of independent WL and TL in $R$ and $R'$ respectively. The certainty relation becomes 
\be \label{cerinv}
S_{{\cal A}_{W}(R)}(\omega,\omega\circ E_{W})+S_{ {\cal A}_{WT}(R')}(\omega,\omega\circ E_{T})=\log Z\;.
\ee
In what follows the objective will be to compute bounds on both entropic order and disorder parameters independently and verify the previous relation.

\subsection{Line operators}
\label{sm}

 WL and TL are defined in the literature as line operators. In this limit, the expectation values always vanish exponentially with a cutoff. Then, we cannot use them to produce non-trivial bounds on the entropic parameters. The situation is somewhat analogous to the concept of local field in QFT, but the distributional character of line operators is more singular. However, line operators have the correct commutation relations for non-local operators and implement the symmetry transformations on the complementary algebra. In this sense, we will find the expression for the TL especially useful for our purposes.  

As it is well known, the WL line operator for a closed path $\Gamma$ is given by
\be
W^{r}_\Gamma=P \,\, \textrm{tr}\, e^{i\, g\, \int_{\Gamma} \, A_\mu \, dx^\mu }\,,
\ee
with $P$ the path ordering of operators, and $A_{\mu}(x)\equiv A_{\mu}^a(x)T^{r}_a$ the gauge potential in a representation $r$ of the group. These line operators are  labeled by group representations. 

 One explicit construction of a non-Abelian TL is described in \cite{Reinhardt_2003}. We begin by fixing the gauge $A_0=0$, and follow the canonical quantization scheme
\begin{equation}
[A^a_i(\vec x),E^b_j(\vec y)]=i \delta_{ij}\;\delta^{ab} \;\delta(\vec x-\vec y)\,.
\end{equation}
In this gauge, first consider the generator of gauge transformations, i.e. the covariant divergence of the electric field $D\cdot E$. This gives zero when applied to any vector in the physical Hilbert space. This is the non-Abelian Gauss constraint
\be
(D\cdot E)^a= \partial_i E_i^a + g f^{abc} A_i^b E_i^c \equiv 0\,,
\ee
where $f_{abc} $ are the structure constants of the gauge group. We can smear such Gauss law operator by a compactly supported Lie algebra valued function $\Lambda(x)=\Lambda^a(x)T^a$, and integrate by parts
\be
\int d^3x\, \textrm{tr} ((D_i \Lambda)\, E_i ) = \int d^3x\, (\partial_i \Lambda^a +g f^{abc} A_i^b \Lambda^c)\, E_i^a
=- \int d^3x\, \textrm{tr} \,\Lambda\, D_i E_i\equiv 0\,. \label{expo}
\ee
This  holds even if $\Lambda$ is not continuous, and its derivatives have delta function contributions. Exponentiating this operator 
\be
e^{i\int d^3x\,  \textrm{tr}( (D \Lambda)\cdot E )} =e^{-i\int d^3x\,  \textrm{tr} (\Lambda D\cdot E )}\equiv 1\,,
\ee
we obtain the identity on the physical Hilbert space. 

We want to construct the TL corresponding to a closed curve $\Gamma'$ which is the boundary of a two-dimensional surface $\Sigma$, $\partial \Sigma=\Gamma'$. The idea in \cite{Reinhardt_2003}
starts by choosing vector fields $j^a_i$, smooth outside $\Gamma'$, with vanishing curl. As they have a vanishing curl they have fixed circulation around any curve $\Gamma$ linked with $\Gamma'$. Let us call $2\pi\omega_{a}^{\vee}$ to this circulation.\footnote{The reason for this notation will be clear later.} The vector fields $j_i^a$ can be written locally as gradients $j_i^a=\partial_i \Lambda^a$ for some functions $\Lambda^a$ but not globally. The functions $\Lambda^a$ can be chosen such that they are smooth outside $\Sigma$ but have a discontinuity $2\pi\omega_{a}^{\vee}$ across $\Sigma$, which gives the non-trivial circulation of $j^a_i$.  Using these vector fields we can now define the following operator
\be
T=e^{i g^{-1}\int d^3x\, (j_i^a +g f^{abc} A_i^b \Lambda^c)\, E_i^a}\,.\label{t2}
\ee
The exponent is equal to (\ref{expo}) except for the replacement of the derivatives of $\Lambda^a$ by $j^a$. This replacement keeps only the regular piece of the gradient of $\Lambda^a$. Since this is a pure gauge transformation except for this non-regular piece, we can equivalently write this operator in terms of the discontinuity $2\pi\omega_{a}^{\vee} $ of $\Lambda^a$ across $\Sigma$
\be
T\equiv  e^{i \frac{2\pi}{g}\,\int_\Sigma d\sigma\, \omega_{a}^{\vee}\; E_i^a\; \eta_i}\,,\label{t1}
\ee
with $\eta_i$ the unit normal to the surface. 
Defined in this way, this operator generically depends on the surface $\Sigma$ used to define $\Lambda$ instead of the curve $\Gamma'$. Besides, it can also be non-gauge invariant. Both of these problems disappear if we choose the discontinuity $2\pi\omega_{a}^{\vee}$ such that
\be 
e^{i 2\pi\;\omega_{a}^{\vee}\; T^a}=z\in Z\;\label{z}
\ee
belongs to the center of the gauge group \cite{Reinhardt_2003}. This defines a TL that we name $T_{\Gamma'}^z$, labeled by $\Gamma'$ and an element $z\in Z$. 

Furthermore, the action of this (non-smeared) TL on WL that encircle the loop is to insert the element $z$ on the path ordered WL on $\Sigma$ along the path of the loop. This is because $T_{\Gamma'}^z$, being an exponential of the conjugated field $E_i$ in (\ref{t1}), just effects a displacement of the gauge connection $A^a\rightarrow A^a+g^{-1}\delta(x-\Sigma)\,2\pi\omega_{a}^{\vee} $. When this displacement is exponentiated we obtain the center group element.  Notice this does not depend on the position of the insertion in the WL because it is a central element. It then escapes the trace used to define the WL as the corresponding character of $z$:
\be
T_{\Gamma'}^z W^r_{\Gamma} (T_{\Gamma'}^z)^{-1}= \frac{\chi_r(z)}{d_r}  W^r_\Gamma\,,
\ee  
were we have assumed $\Gamma, \Gamma'$ are simply laced to each other. Here $d_r$ is the dimension of the representation $r$, and $\frac{\chi_r(z)}{d_r}$ is in fact a character of the center $Z$. This  indicates the non local class to which the WL belongs. 
On the other hand, it is immediately clear from the definition that $T_{\Gamma'}^z$ commutes with gauge invariant operators with support in any simply connected region outside $\Gamma'$. The reason is that in such regions, $T_{\Gamma'}^z$ acts as a simple gauge transformation.

It may seem that in this definition of line operators there is a certain asymmetry between WL and TL. While WL are labeled by representations of the gauge group, TL are labeled by elements of the center. However, TL labeled by representations of the dual group can be defined \cite{Kapustin_2006}. If the theory can be formulated with a dual gauge field, these TL are just the WL of such a dual field. In general, however, these are not line operators but surface operators in the present context. Note also that only the class of the WL, determined by the representation of the center, is relevant to the present problem. Below we will end up with a fully symmetric approach to smeared WL and TL.

For our purposes, the next natural step would be to find smeared versions of these line operators.   It turns out that the problem of finding a good definition of smeared gauge invariant WL and TL in the non-Abelian set-up is far from trivial.  Certain efforts in that direction can be found in \cite{Narayanan_2006,Lohmayer:2011si}. However, these results require renormalization to take the continuum limit. Further, they are thought for Euclidean smeared loops, and the smearing functions do not have compact support.  The problem of constructing gauge invariant smeared loops with bounded support in real-time has not been considered in the literature to our knowledge.

 In the next sub-sections, we will study this problem from a different perspective, using an enlarged Hilbert space of non-gauge invariant wave functions. This will allow a simple non-perturbative definition suited to the relative entropy computation. Since we are mainly interested in the weak coupling limit, consisting of an orbifold of Maxwell fields, it is natural to seek also for particular smeared operators that would perform the task in this particular limit. In appendix \ref{weak} we briefly consider some simple ideas on gauge invariant smeared operators in the weak coupling limit, and look more closely at the encountered difficulties. As the results appear unavailing, the uninterested reader may as well skip such discussion. 
 
As a final remark, the previous standard construction of line operators naively suggests that there should be smeared WL for any representations of the gauge group. However, this idea does not seem to have any clear justification since there is no way to measure such representation.  The WL will be an element of the class of non-local operators of the ring (that is, associated with a representation of the center of the group). But crucially, it cannot carry any specific representation under the gauge group itself. The construction of smeared operators we present below somehow makes this observation more acute and transparent since the Wilson loops will not be labeled by representations, but by the weights of the group.
 
 \subsection{Gauge non invariant Hilbert space}
\label{another}

The natural route to obtain bounds on the relative entropies would be to find an explicit form of the corresponding gauge invariant smeared non-local WL and TL, and compute their expectation values in the weak coupling limit. However, as we have already stressed above, an explicit expression of smeared WL seems to be difficult to obtain in a closed-form. However, the relative entropy we want to evaluate is a very flexible tool, that allows us to evaluate bounds without having such an explicit expression.  In the upcoming sections, we will show how to compute bounds on the relative entropy order parameters using an enlarged Hilbert space description.

Our construction of smeared WL and TL in a larger,  non-gauge invariant Hilbert space is non-perturbative. Further, using this construction we will be able to obtain expectation values in the perturbative regime. Perturbative corrections might be computed in principle using the same ideas.     
Working in the non gauge invariant Hilbert space allows us to recast the  
calculation of the order parameter in a way that is very similar to the Maxwell $Z_n$ model of section \ref{entropic}, or simple generalizations. 

Our first task is to write the relative entropy we seek to compute as one in a non gauge invariant Hilbert space. To be clear, it is better to describe this step in the lattice, so we start briefly describing the lattice setup. 
   
\subsubsection*{Gauge invariant and non invariant algebras}   
   
In the lattice, gauge theories are described by oriented link variables $U_l\in G$, for each link $l$, and  gauge transformations $g_a$ associated to vertices $a\in V$. The link with opposite orientation $\bar{l}$ has associated the inverse group element $U_l^{-1}$. The non-gauge invariant Hilbert space ${\cal H}$ is given by  arbitrary wave functions $\Psi[U_l]$ of the link variable configurations. The scalar product is 
\be  
\langle \Psi_1|\Psi_2\rangle= \int \prod_l dU_l\, \Psi_1[U_l]^* \,\Psi_2[U_l]\,,
\ee
with the usual Haar measure on the group. 
The set of operators ${\cal B}({\cal H})$ in ${\cal H}$ admits a set of local generators based on links. We have ``coordinate'' or ``magnetic''  variables defined by the operators $D_{r, ij}^{(l)}$, associated with the link $l$ on the lattice. They just multiply the wave function by the numerical value of the matrix entry $ij$ of  the irreducible representation $r$ of the gauge group, evaluated on the link variable $U_l\in G$:
\be
D_{r, ij}^{(l)}\Psi[U_1,\cdots, U_l, \cdots, U_N]= D_{r, ij}(U_l)\,\Psi[U_1,\cdots, U_l, \cdots, U_N]\,.
\ee
 ``Momentum'' or ``electric'' variables are given by operators $L^{(l)}_g$ such that
\be
L^{(l)}_g \Psi[U_1,\cdots, U_l, \cdots, U_N]=\Psi[U_1,\cdots, g U_l, \cdots, U_N]\,.
\ee 
These local generators allow us to define local algebras ${\cal B}(R)$ in ${\cal H}$ for any subset of links $R$ in the lattice. 

Gauge transformations $g_a$ act locally on the vertices of the lattice as
\be
C_{a,g}= \prod_{l=(ab)} L_g^{(l)}\,. 
\ee
 The physical Hilbert space ${\cal F}\subset {\cal H}$ is formed by the gauge invariant vectors $C_{a,g} |\psi\rangle= |\psi\rangle$  for any vertex $a$ and group element $g$.  The algebra of gauge invariant operators is likewise defined by the operators satisfying $C_{a,g} {\cal O} C_{a,g}^{-1}={\cal O}$. 

We are interested exclusively in the operators in the region $R$. We define the gauge invariant algebra ${\cal A}_R$ as the one formed by gauge invariant elements in ${\cal B}_R$. We can express the relation between the two algebras using a conditional expectation $E_G$ produced by the gauge transformations 
\be
E_G(x)=\prod_{a\in V}  \int d g_a\,  C_{a,g_a}\, x\,  C_{a,g_a}^{-1}\,,
\ee 
where the Haar measure is normalized to one. This conditional expectation acts locally and projects to the gauge invariant operators
\be
E_G({\cal B}(R))= {\cal A}_{W}(R)\,.
\ee
However, notice that ${\cal A}_{W}(R)$ contains non-local WL operators as well as local ones. 

In general, the conditional expectation $E_G$ does not kill non gauge invariant operators but picks up their gauge invariant support.  As an example, a gauge-invariant WL can be produced by the non gauge invariant operator  
\be 
\mathcal{O}_\Gamma=D_{r,i_1j_1}^{(l_1)}\cdots D_{r,i_n j_n}^{(l_n)}\;,\label{dsa}
\ee
where it is understood that the link $l_k$ shares a vertex with the link $l_{k+1}$, including a shared vertex between $l_1$ and $l_n$, forming a closed path $\Gamma$. The operator ${\cal O}$ is obviously not gauge invariant. To see if it has some gauge invariant support we need to average over the gauge transformations. We have to average independently over gauge transformations based at each vertex. Recall that if the link $l=(ab)$ goes from vertex $a$ to vertex $b$, then the index $i$ of the operator $D_{r,ij}^{(l)}$ transforms in representation $r$ under gauge transformations based on $a$ and the index $j$ in representation $r^*$ under gauge transformations based on $b$. Averaging over gauge transformations on a vertex $a$ shared by the links $l_k, l_{k+1}$ corresponds to the transformation  
\bea\label{globex}
D_{r,i_k j_k}^{(l_k)}D_{r,i_{k+1} j_{k+1}}^{(l_{k+1})}\rightarrow \sum_{s,t} D_{r,i_k s}^{(l_k)}\left(\int dg\,   D_{r,s j_k}(g) D_{r,i_{k+1}t}^{*}(g)\right) D_{r,t j_{k+1}}^{(l_{k+1})}
\nonumber\\ = \delta_{j_k, i_{k+1}} \sum_s D_{r,i_k s}^{(l_k)}D_{r,s j_{k+1}}^{(l_{k+1})}
\;,
\eea
where we have used the orthogonality relations of irreducible representation matrices.  
Averaging over all vertices we arrive at
\be \label{gaue}
E_{G}(\mathcal{O}_\Gamma)=\left(\prod_k \delta_{j_k i_{k+1}}\right)\; W^r_\Gamma\,.
\ee
We can thus arrive at the Wilson loop $W^r_\Gamma$ just by doing appropriate projections over non-gauge invariant operators. Note that the averaged operator vanishes if the string of link operators is open, or if the indices do not match, but it is certainly not necessary to sum over the indices in (\ref{dsa}).

To show a particular operator with non trivial gauge invariant support  we can write $U^{(l)}=e^{i \,g\, \epsilon \,A_l^a T_a}$, defining the gauge potential, where $\epsilon$ is the lattice spacing, and $D_{r,ij}^{(l)}=\left(e^{i \,g\, \epsilon \,A_l^a T^r_a}\right)_{ij}$. 
Calling $F^a_{ij}$ to the $T^a$ matrices for the fundamental representation $F$, in the continuum limit $\epsilon\rightarrow 0$ we have
\be
O^{1}_l=\textrm{tr} \left((1/d_F+2\, F^1) \, D_F^{(l)} \right)\sim 1+ i g \, \epsilon A^a_l \, 2\, F^a_{ij} F^1_{ji}=1+ i g \, \epsilon A^1_l\sim e^{i g \, \epsilon A^1_l}\,. 
\ee
Taking products of these $O^1_l$ along a path $\Gamma$ 
\be
O_\Gamma^1=\prod_{l\in\Gamma} O^1_l\,,
\ee
we would get a lattice representation approaching the gauge non-invariant line operator $e^{i\int_\Gamma dx^i A^1_i}$ in the continuum. In the lattice,  once averaged over the gauge group, $O^1_\Gamma$ is proportional to the fundamental Wilson loop. Therefore it has non-zero gauge invariant support.   

\subsection{Relative entropies in the gauge non invariant space}

Let us start considering the WL order parameter. We want to compute, for example,  the relative entropy between the gauge invariant states $\omega$ and $\omega\circ E_W$ in the ring algebra ${\cal A}_{W}(R)$. These states can be lifted to gauge invariant states in the full algebra ${\cal B}(R)$ on ${\cal H}$ as $\omega\circ E_{G}$ and $\omega\circ E_W\circ E_{G}$. Because of the conditional expectation property \cite{petz2007quantum}, we have the equality of relative entropies for invariant states\footnote{In the early discussions about entanglement entropy in gauge theories the problem was raised on how to cut tensor product of local Hilbert spaces for some region $R$ and its complement. Some of the proposals used gauge non-invariant Hilbert spaces \cite{Ghosh:2015iwa,Soni:2015yga}. A more transparent setup indicates that one has to chose an algebra for $R$ rather than a Hilbert space partition as a tensor product. The ambiguities of assigning algebras to regions in a lattice are more general and do not have to do specifically with gauge theories. Different algebras with the same physical content in the continuum limit give place to the same relative entropy quantities \cite{Casini:2013rba}. Then this type of ultraviolet ambiguities associated with the boundaries will not bother us here. }
\be
S_{{\cal A}_{W}(R)}(\omega|\omega\circ E_W) =S_{{\cal B}(R)}(\omega\circ E_{G}|\omega\circ E_W\circ E_{G})\, . \label{oop}
\ee
Notice that the state $\omega\circ E_{G}$ in the gauge non invariant algebra is the usual vacuum state computed by path integration. We will therefore also call $\omega$ to the gauge invariant vacuum state when considered in ${\cal B}(R)$.   
We want to give a different expression to the second state on the rhs of this equation.  

The TL is constructed in the lattice by taking products of link electric operators. Given a two dimensional surface $\Sigma$ we define \cite{Casini:2020rgj}
\be
T^z_\Sigma= \prod_{l\perp \Sigma} L^{(l)}_z\,,\label{lop}
\ee
for an element $z\in Z$, the center of the gauge group. These operators have the same multiplication law as the elements $z$ of the center of the group. Because $z$ is a central element this operator commutes with all gauge transformations. It is gauge invariant
\be
C_{a,g}\, T^z_\Sigma \,C_{a,g}^{-1}=T^z_\Sigma \rightarrow E_G(T^z_\Sigma)= T^z_\Sigma\,.
\ee
Further, in the gauge invariant Hilbert space it is also independent of the surface $\Sigma$ provided the boundary $\partial \Sigma$ is kept fixed. The reason is that for such different $\Sigma, \Sigma'$, the difference is just a gauge transformation. More generally, given a general gauge transformation
\be
{\cal G} =\prod_{a\in V} C_{a,g_a}\,,
\ee 
 the action of the 't Hooft loop $T^z_\Sigma$ as a unitary transformation  on the gauge invariant algebra is equivalent to any other representative operator
\be
\tilde{T}^z= T^z_\Sigma\, {\cal G}={\cal G}\, T^z_\Sigma\,.\label{psa}
\ee 
It follows from this equation that general TL representatives $\tilde{T}^z$ are non gauge invariant. However, for a general gauge transformation ${\cal G}_1$ we have another one ${\cal G}_2$ such that 
\be
\tilde{T}^z {\cal G}_1= {\cal G}_2 \tilde{T}^z\,. \label{tertius}
\ee

Now we extend the conditional expectation~(\ref{condW}) described above to the  non gauge invariant algebra ${\cal B}(R)$. We call this extension $E_{\tilde{W}}$. It is defined as
\be
E_{\tilde{W}}(x)= \sum_{z\in Z} \tilde{T}^z \,x\,  (\tilde{T}^z)^{-1}\;.\label{lopa}
\ee
It is direct to show that,
\be\label{pok0}
E_{W}\circ E_G=E_{\tilde{W}}\circ E_G \,. 
\ee
This extension of course depends on the group of representatives we have chosen. However, because of property (\ref{tertius}) we have
\be\label{pok1}
E_{\tilde{W}}\circ E_G= E_G\circ E_{\tilde{W}} \,. 
\ee
We summarize these operations in the commutative diagram,
\bea
{\cal B}(R) & \overset{E_{\tilde W}}{\longrightarrow} & \tilde {\cal B}_{\rm add}(R)\nonumber \\
\downarrow E_G\! &  & \:\downarrow E_G\\
\mathcal{A}_{W}(R) & \overset{E_W}{\longrightarrow} & \mathcal{A}_{\rm add}(R)\;,\nonumber 
\eea
where $\tilde {\cal B}_{\rm add}(R)$ corresponds to the elements invariant under $E_{\tilde W}$.

Relations \eqref{pok0} and \eqref{pok1} allow us to recast the relative entropy in the gauge non-invariant space (\ref{oop}) in a useful form
\be \label{pok}
S_{{\cal A}_{W}(R)}(\omega|\omega\circ E_W) =S_{{\cal B}(R)}(\omega\circ E_{G}|\omega\circ E_W\circ E_{G})=S_{{\cal B}(R)}(\omega|\omega\circ E_{\tilde{W}})\, , 
\ee
where the gauge invariant state $\omega$ is obtained by path integration.  
  
Let us see in more detail the action of the non invariant TL on ${\cal B}(R)$. The operator $T^z_\Sigma$ does not commute only with the operators $D^{(l)}_{r,i,j}$ where $l\perp \Sigma$. For such operator the unitary action of  $T^z_\Sigma$  gives a phase factor $\chi_{z^*}(z)$, where $z^*$ is the representation induced by $r$ on the center $Z$. Therefore, operators in ${\cal B}(R)$ can be decomposed into classes 
\be
{\cal B}(R)=\oplus_{z^*}  {\cal B}_{z^*}(R)\,,\hspace{1cm} x\in {\cal B}_{z^*}(R)\iff T^z_\Sigma \,x= \chi_{z^*}(z) \, x  \,T^z_\Sigma\,.
\ee
Some operators in each class are annihilated by $E_G$, but the ones that are not annihilated must form non-trivial classes after averaging. This is due to the bimodule property satisfied by conditional expectations, which for the previous discussion implies
\be 
T^z_\Sigma \,x= \chi_{z^*}(z) \, x  \,T^z_\Sigma\,\Longrightarrow\, E_G (T^z_\Sigma \,x)=E_G ( \chi_{z^*}(z) \, x  \,T^z_\Sigma))\,\Longrightarrow\, T^z_\Sigma \,E_G ( x)= \chi_{z^*}(z) \, E_G ( x)  \,T^z_\Sigma\;.
\ee
Then let
\be
 \tilde{W}_{z^*} \in {\cal B}_{z^*}(R)\,, \hspace{1cm} E_G( \tilde{W}_{z^*})=W_{z^*}\,,
\ee    
with $W_{z^*}$ a gauge invariant operator of class $z^*$. 
We have necessarily
\be
 T^z_\Sigma \tilde{W}_{z^*} (T^z_\Sigma)^{-1}=\chi_{z^*}(z)  \tilde{W}_{z^*}\,, \hspace{1cm} 
 T^z_\Sigma W_{z^*} (T^z_\Sigma)^{-1}=\chi_{z^*}(z)  W_{z^*}\,.
\ee  
If we consider  any other representative of the TL as in (\ref{psa}) we get instead
\be
\tilde{T}^z \tilde{W}_{z^*} (\tilde{T}^z)^{-1}=\chi_{z^*}(z) \, {\cal G}\, \tilde{W}_{z^*} \,{\cal G}^{-1}\,, \hspace{1cm} 
\tilde{T}^z W_{z^*} (\tilde{T}^z)^{-1}=\chi_{z^*}(z)  W_{z^*}\,. \label{derecha}
\ee
Thus, on the non gauge invariant candidate WL the generic TL introduces both the usual phase and a gauge transformation. This will be important for the applications we have in mind.  

 To understand physically why the equality between relative entropies in (\ref{pok}) holds, note that non-gauge invariant operators which do not commute with $T_\Sigma$ are either killed by $E_G$ or produce a non-trivial class under averaging. Hence, there are many operators in the larger algebra ${\cal B}(R)$ which are non-invariant under the conditional expectation $E_{\tilde{T}}$ but do not increase the distinguishability between states in the relative entropy because they have zero expectation value. The existence of these ``superfluous'' operators is the origin of the conditional expectation property (\ref{oop}).\footnote{In fact, in the present representation, the gauge invariant algebra also contains many trivial operators spanned by the gauge transformations $C_{a,g}$. In particular, the difference between $T_\Sigma^z$ and $T_{\Sigma'}^z$ is a combination of these operators. The gauge constraints $C_{a,g}$ are set to eigenvalue $1$ on the gauge invariant Hilbert space.  
 Therefore they do not add to the distinguishability in the relative entropy between gauge invariant states, and could as well be neglected or assimilated to the identity operator. At a more technical level, the relative entropy can be computed from the GNS representation of the states and this representation already takes these operators to the identity.}

\subsection{Smeared non invariant operators in the continuum}

To treat the problem in the continuum we fix the gauge $A^0=0$ and $[A_i(x),E_j(y)]=i \delta_{i,j} \delta^3(x-y)$ in the non gauge invariant Hilbert space.  We start from the expression (\ref{t2}) for the TL. We fix a function $\Lambda$  with a smooth gradient $j_i=\partial_i \Lambda$. It does not depend on the Lie algebra indices. We fix the circulation of $j$ to be $1$, such that $\Lambda$ has a discontinuity $1$ across the surface $\Sigma$ with boundary $\Gamma'$. We take $\Sigma$ to cut a cross section of the ring $R$ were our WL are seated.  This $j_i$ is the $\tilde{j}_i$ of (\ref{thooftloop}) inside $R$, having zero curl and charge equal to $1$. Using this standard $\Lambda$ the TL expression gets 
\be
\tilde{T}_{\omega^{\vee}}=e^{i \frac{2\pi}{g}\int d^3x\, (j_i\, \omega_a^{\vee} \,+\,g \,f_{ab}^{c} \,A_i^b\, \Lambda \,\omega_c^{\vee})\, E_i^a}\,,\label{t22}
\ee
corresponding to the element of the center $z=e^{i 2\pi\omega_{a}^{\vee} T_a}$. Notice we now have labeled the TL with the vector $\omega^{\vee}$ instead of $z$ because there are multiple $\omega^{\vee}$ giving the same $z$, and they are different operators in the non gauge invariant space.  

It will be convenient to choose the vectors $\omega^{\vee}$ for the different $z\in Z$  so that they belong to the Cartan subalgebra. This is a maximal set of commuting Lie algebra elements and the reason for this choice will be clarified in the next section. In this case, for any two vectors $\omega^{\vee}, \tilde{\omega}^{\vee}$
\be
[\omega^{\vee}_a \,T^a,\tilde{\omega}^{\vee}_b \, T^b]=0\,\,\Longrightarrow\,\, f^{abc}\; \omega^{\vee}_a \; \tilde{\omega}^{\vee}_b =0 \,,   
\ee
where the structure constants $f^{abc}$ are completely antisymmetric. 
This implies 
\be
 \left[(j_i \,\omega^{\vee}_a +g \,f^{c}_{ab}\,A_i^b \,\Lambda \,\omega_c^{\vee})\, E_i^a,(j_i\, \tilde{\omega}_a^{\vee} +g \,f^{c}_{ab}\, A_i^b\, \Lambda \,\tilde{\omega}_c^{\vee})\, E_i^a\right]=0\,,
\ee
and the TL commute with each other.

Now we construct non gauge invariant WL which we can use as a subalgebra to evaluate a lower bound in the relative entropy. In this construction we are guided by the following considerations:
\begin{itemize}
\item We want to keep the algebra of the WL as compact as possible, and then require that the action of the TL maps the WL in themselves. That is, we ask the additional gauge transformation in (\ref{derecha}) to be the identity. 

\item In consequence, to have a non-trivial element after averaging, eq. (\ref{derecha}) implies the commutation relation with the TL must be fixed by the phase factors $\chi_{z^*}(z)$ alone. 

\item In the action of the TL on the WL, it will be convenient that the discontinuity in the exponent of the TL does not play any role. This guarantees that we can understand the algebra in the continuum, without recurring to a lattice regulator. 
\end{itemize}

The first and the third conditions are immediately satisfied if we choose our non gauge invariant WL to be of the form
\be
\tilde{W}_\omega= e^{i\, g \, \int J \cdot A^a \omega_a}\,,\label{lan}
\ee
where we choose  $J$ of unit charge, and $\omega_a$ is a vector whose index runs again over the Cartan subalgebra. In fact, we get
\be
\tilde{T}^{\omega^{\vee}} \,\tilde{W}^{\omega} \,  (\tilde{T}^{\omega^{\vee}})^{-1}= e^{i g \int J_i (A_i^a +\frac{2\pi}{g} j_i \omega^{\vee}_a +2\pi\, f^{abc} A^b_i \Lambda \omega^{\vee}_c  )\, \omega_a}=  e^{i\,2\pi\, \omega^{\vee}_a \omega_a}\,  \tilde{W}^\omega\,.
\ee
 Then, to satisfy the second condition, we are bounded to choose $\omega$ and $\omega^{\vee}$ such that $e^{i\,2\pi\, \omega\cdot\omega^{\vee}}$ gives an appropriate character of the representations of $Z$.

It remains to smear the TL. This is easily done since we only have to deform it around the boundary of $\Sigma$ without altering the commutations relations with the WL.  We would start from the expression (\ref{t22}) and insert a smearing function $h(x)$ that goes to zero at $\Gamma'$ and saturates to $1$ outside the ring $R'$. We then define
\be
\tilde{T}^{\omega^{\vee}} =e^{i \frac{2\pi}{g}\,\int d^3x\, h(x) (j_i \,\omega^{\vee}_a +g \,f^{abc}\, A_i^b \,\Lambda\,\omega^{\vee}_c)\, E_i^a}\,.\label{t222}
\ee
This has the correct commutation relations with loops and local operators outside $R'$. 
We could use this expression in a non perturbative setup, but in the weak coupling limit this simplifies. To the lowest order in $g$ we can neglect the second term in the exponent of (\ref{t2}), and write more simply 
\be
\tilde{T}^{\omega^{\vee}} =e^{i \frac{2\pi}{g}\,\int d^3x\,  j_i \,\omega_{a}^{\vee} \, E_i^a}\,,\label{t2222}
\ee
where now $\nabla\times j=\tilde{J}$ is a conserved current that vanishes outside the ring $R'$, and has unit charge. The operators (\ref{lan}) and (\ref{t2222}) generalize the  expressions we have used for the Maxwell field. 

Notice that these smeared TL in the non gauge invariant space does not satisfy the group $Z$ operation anymore. With this statement, we mean that, for example, in $SU(2)$ the only non-trivial TL in the $\mathbb{Z}_2$ algebra is expected to meet $T^2=1$. However, it is easy to see that our definitions \eqref{t222} will not meet this condition. In analogy with the treatment of the case of the Maxwell field above, we can build a closed algebra by taking all integer powers of the smallest charge TL. The conditional expectation will eliminate all non-local operators. The important conditions to meet in our construction are for the algebra to be closed and to contain adequate non-local operators. If the algebra happens to contain any extra additive operators, as is our case above, this can only improve the bounds. For example, for the case of $\mathbb{Z}_2$, only the odd powers of the TL will be non-local, corresponding to only one class of non-trivial non-local operators.

\subsection{Weight and co-weight lattices of Wilson and 't Hooft loops}

In the previous sections we have seen how to define Wilson and 't Hooft loops in the gauge non-invariant Hilbert space. In particular, we described how the commutation relations are preserved under the projection to the gauge invariant part. Parallelly, as described previously, we know in general grounds the algebra between invariant loops in gauge theories
\be
T_z \, W_{z^*}= \chi_{z^*}(z) \, W_{z^*} \, T_z\,,
\ee
where $z$ is an element of the center and $z^*$ is a representation of the center. The commutation relations between the non-invariant smeared loops also involve just a phase, namely $e^{i\,2\pi\, \omega^{\vee}_a \omega_a}$, where the $\omega_a$ and $\omega^{\vee}_a$ are the charge vectors defining the non-invariant 't Hooft and Wilson loops respectively. Although in the non-invariant Hilbert space any phase is possible, the projection to the gauge invariant part will necessarily kill any set of line operators which do not commute as they should. We thus have to find the allowed set of charge vectors $\omega_a$ and $\omega^{\vee}_a$ satisfying
\be 
e^{i\,2\pi\, \omega^{\vee}_a \omega_a}=\chi_{z^*}(z)\;,
\ee
for some representation of the center $z^*$ and some element of the center $z$. To solve this problem, in what follows we assume certain technical knowledge of the representation theory of Lie groups. We have collected all the necessary ingredients in a brief group theory review in appendix \ref{Lie}. If the reader is not familiar with the concepts of weight, co-weight, root, and co-root lattices, the fundamental weights, co-weights, roots, and co-roots, together with the equations relating them we recommend reading such appendix first.

Let us start with the charge vector  $\omega^{\vee}_a$ associated with the 't Hooft loop. It was defined as usual, so that in the abstract group it satisfies
\be 
e^{i\,2\pi\, \omega^{\vee}_a T_a}=z\;,
\ee
where $z$ is an element of the center. Because the center is a group, the possible $\omega^{\vee}$ satisfying this equation for some $z\in Z$ will form a lattice. As mentioned above, it is convenient to choose the first $T_a, a=1,\cdots, l$, with $l$ the rank of the group, to be a basis of the Cartan subalgebra. Therefore $\omega^{\vee}_a$ will be a vector of length $l$.

From now on we will talk only about elements of the Cartan subalgebra of the Lie algebra. As it is a commutative set we can talk about the common eigenvectors in any given representation.    
The eigenvectors in the adjoint representation with non-zero eigenvalues are the ``roots'' $E_{\pm\alpha}$ with eigenvalue $\pm\alpha$, where the $\alpha$'s are vectors of dimension equal to the rank $l$ of the Lie algebra (that is, they are a list of the eigenvalues for the different elements of the Cartan subalgebra). 

Since the center is represented trivially in the adjoint representation we obtain
\be \label{toeq}
e^{i \,2\pi\, \omega^{\vee}\cdot\alpha}=1\rightarrow \, \omega^{\vee}\cdot\alpha = \mathds{Z}\;.
\ee 
where the dot represents the usual inner product of an $l$-dimensional Hilbert space. This equation should be true for any vector in the root lattice. To solve this equation we notice that in any Lie algebra, the weights and roots satisfy
\be 
\frac{2 \alpha\cdot\omega}{\vert\alpha\vert^2}\in \mathds{Z}\;.
\ee
But root systems come in pairs \cite{Goddard:1976qe}. For any root system and root lattice, generated by $l$ fundamental roots $\alpha^{(i)}$, where $i=1,\cdots ,l$, we have a dual one whose fundamental ``co-roots'' are defined by
\be 
\alpha^{\vee}_{(i)}\equiv \frac{2 \alpha^{(i)}}{\vert\alpha^{(i)}\vert^2}\;.
\ee
The relation between roots and co-roots is symmetric, in the sense that we also have
\be 
\alpha_{(i)}\equiv \frac{2 \alpha^{\vee}_{(i)}}{\vert\alpha^{\vee}_{(i)}\vert^2}\;.
\ee 
These co-roots are the roots of the dual group. Together with the dual ``magnetic co-weights'' $\omega^{\vee}$, they also satisfy
\be 
\frac{2 \alpha^{\vee}\cdot\omega^{\vee}}{\vert\alpha^{\vee}\vert^2}\in \mathds{Z}\;,
\ee
and so
\be 
\frac{2 \alpha^{\vee}\cdot\omega^{\vee}}{\vert\alpha^{\vee}\vert^2}=\alpha\cdot\omega^{\vee}\in \mathds{Z}\;.
\ee
Comparing this equation with~(\ref{toeq}), we conclude that the charge vector defining the 't Hooft loop needs to be of the form
\be 
2\pi\,\omega^{\vee}\;,
\ee
with $\omega^{\vee}$ any vector in the co-weight lattice. Finally, note that given a vector $\omega^{\vee}$ in the co-weight lattice, there is a single element of the center related to it by $e^{i2\pi\,\omega^{\vee}\cdot T}=z$. The co-weight lattice then decomposes into classes labeled by elements $z$ of the center of the gauge group. Each class $z$ is just defined as the set of points in the co-weight lattice associated with $z$. This is the known fact that the quotient between the lattice of co-weights $\Lambda_{\omega^{\vee}}$ and the lattice of co-roots $\Lambda_{\alpha^{\vee}}$ is equivalent to the center of the group, namely
\be 
\Lambda_{\omega^{\vee}}/\Lambda_{\alpha^{\vee}}\sim Z \;.
\ee

Having understood this we now move to the charge vector $\omega$ defining the Wilson loop. It has to satisfy
\be 
e^{i2\pi \omega\cdot\omega^{\vee}}=\chi_{z^*}(z)\;,
\ee
for some representation of the center $z^*$, and where now $z$ is the element canonically associated with $\omega^{\vee}$.

Given the solution for the $\omega^{\vee}$'s, a natural guess for the $\omega$'s is that they are vectors in the weight lattice $\Lambda_{\omega}$. This guess turns out to be correct. We just need to prove this is true for the fundamental weights $\omega^{(i)}$, where $i=1,\cdots ,l$. These are the weights that generate the weight lattice. We thus need to understand the inner product $\omega^{(i)}\cdot \omega^{\vee (j)}$. To do this, we remind that the fundamental weights are related to the fundamental roots through the Cartan matrix $A_{ij}$, defined as
\be 
A_{ij}=\frac{2 \alpha^{(i)}\cdot\alpha^{(j)}}{\vert\alpha^{(j)}\vert^2}\;,
\ee
see appendix \ref{Lie}. Using this matrix the precise relation is
\be 
\omega^{(i)}=\sum_{j}A^{-1}_{ji}\alpha^{(j)}\;.
\ee
The inner product then becomes
\be 
\omega^{(k)}\cdot \omega^{\vee (l)}=\sum_{ji}A^{-1}_{jk}A^{-1}_{il}\alpha^{(j)}\cdot\alpha^{\vee (i)}=\sum_{ji}A^{-1}_{jk}A^{-1}_{il}A_{ji}=A^{-1}_{lk}=A^{-1}_{kl}\;.
\ee
We conclude that for a fundamental weight $\omega^{(i)}$ defining the Wilson loop (canonically associated with a representation $z^{*}$ of the center of the group) and a fundamental co-weight $\omega^{\vee (j)}$ defining the t' Hooft loop (canonically associated with an element $z$ of the center) the phase appearing in the commutator is
\be 
e^{i2\pi \omega^{(i)}\cdot\omega^{\vee (j)}}=e^{i2\pi A^{-1}_{ij}}\;.
\ee
Going now to the tables of Lie groups we find that indeed
\be\label{magicrel} 
e^{i2\pi A^{-1}_{ij}}=\chi_{z^*}(z)\;.
\ee
To prove this in more detail, and clarify why and how the indices in both sides match, we step on the known quotient between weight and root lattices, given by the center
\be 
\Lambda_{\omega}/\Lambda_{\alpha}\sim Z^*\;.
\ee
This means we can define one class per element $z^{*}\in Z^*$, namely
\be 
[\omega_{z^*}]\equiv [\omega_{z^*}+\sum_{i=1}^l n_i \alpha^{(i)}]\;,
\ee
where $\omega_{z^*}$ is any representative of the class and $n_i \in \mathds{Z}$. As for these representatives, we can take the fundamental weights. We just need to take care because there are $l$ fundamental weights, where $l$ is the rank, while there are $Z-1$ non-trivial elements $z^{*}\in\Lambda_{Z}$. The identification is not always one-to-one (in $SU(N)$ it is one-to-one) since different fundamental weights might be connected by the root lattice.\footnote{An extreme example of this is the gauge group $E_8$. This group has rank $8$, so $8$ fundamental weights, but no center. What happens is that the $8$ fundamental weights are also the $8$ fundamental roots.} We just need to choose a subgroup of fundamental weights that is linearly independent modulo the root lattice. This provides a many to one map $\omega^{(i)}\rightarrow\omega_{z^*}$.

Similarly (or dually), for the co-weight and co-root lattices we have
\be 
\Lambda_{\omega^{\vee}}/\Lambda_{\alpha^{\vee}}\sim Z\;.
\ee
and we can define classes labeled by $z\in Z$, namely
\be 
[\omega_{z}^{\vee}]\equiv [\omega_{z}^{\vee}+\sum_{i=1}^l n_i \alpha_{(i)}^{\vee}]\;.
\ee
Again, we have generically a many to one map from fundamental co-weights to representatives of the center. We just need to choose a subgroup of fundamental co-weights which is linearly independent modulo the co-root lattice.

Given these observations, for any class $z$ we can now construct a representation of $Z^*$. This is defined as
\be 
[\omega_{z}^{\vee}]\rightarrow e^{i2\pi [\omega_{z^*}]\cdot [\omega_{z}^{\vee}]}\;.
\ee
Notice this is well defined because we have the relations, see App~\ref{rept}
\bea
\omega\cdot\alpha^{\vee}&=&\mathds{Z}\nonumber\\
\omega^{\vee}\cdot\alpha &=&\mathds{Z}\;.
\eea
Notice also that
\be 
e^{i2\pi [\omega_{z_1^*}]\cdot [\omega_{z}^{\vee}]}e^{i2\pi [\omega_{z_2^*}]\cdot [\omega_{z}^{\vee}]}=e^{i2\pi ([\omega_{z_1^*}]+ [\omega_{z_2^*}])\cdot [\omega_{z}^{\vee}]}=e^{i2\pi [\omega_{z_1^*z_2^*}]\cdot [\omega_{z}^{\vee}]}\;,
\ee
so this is indeed the representation $z$ of $Z^*$. There are $\vert Z\vert$ of them, as expected. Since the group is abelian, the representations are one-dimensional and those numbers as just the characters. But since the representatives are fundamental weights we obtain
\be 
\chi_{z^*}(z)=e^{i2\pi [\omega_{z^*}]\cdot [\omega_{z}^{\vee}]}=e^{i2\pi \omega^{(i)}\cdot\omega^{\vee (j)}}=e^{i2\pi A^{-1}_{ij}}\;.
\ee
We now present two simple examples of this abstract discussion. For $SU(2)$ the rank is one. The fundamental weights and roots are
\bea 
\alpha = 1&\rightarrow & \frac{2\alpha\cdot\omega}{\vert\alpha\vert^2}=1\rightarrow\omega=\frac 12 \,,\nonumber\\
\alpha^{\vee} = 2&\rightarrow &\frac{2\alpha^{\vee}\cdot\omega^{\vee}}{\vert\alpha^{\vee}\vert^2}=1\rightarrow\omega^{\vee}=1\,.
\eea
The Cartan matrix is $A=2$ and therefore $A^{-1}=1/2$. We have
\be 
T_{\omega^{\vee}}W_{\omega}T_{\omega^{\vee}}^{-1}=-W_{\omega}\;.
\ee
For $SU(3)$ the rank is $2$. The fundamental weights and roots are
\bea 
&\alpha_1= &(1,0)\,\,\,\,\,\,\,\,\,\,\,\,\,\,\,\,\,\,\,\,\omega_1=\frac 16 (3,\sqrt 3)\,,\nonumber\\
&\alpha_2= &\frac{1}{2}(-1,\sqrt{3})\,\,\,\,\,\,\,\,\,\,\,\,\,\,\,\,\,\,\,\omega_2=\frac{1}{6}(0,2\sqrt{3})\,,\nonumber\\
&\alpha_1^{\vee}=&(2,0)\,\,\,\,\,\,\,\,\,\,\,\,\,\,\,\,\,\,\,\,\,\,\,\,\,\,\,\,\,\,\,\,\,\omega_1^{\vee}=\frac{1}{\sqrt{3}}(\sqrt 3,1)\,,\nonumber\\
&\alpha_2^{\vee}=&\frac{1}{\sqrt{3}}(-\sqrt 3,1)\,\,\,\,\,\,\,\,\,\,\,\,\,\,\,\,\,\,\,\,\,\,\omega_2^{\vee}=\frac{1}{\sqrt{3}}(0,2)\;,
\eea
and the Cartan matrix is
\begin{equation}\label{GellMann}
A=
\begin{pmatrix}
2 & -1 \\
-1 & 2 
\end{pmatrix}\,,
\qquad\qquad
A^{-1}=
\frac{1}{3}\begin{pmatrix}
2 & 1 \\
1 & 2 
\end{pmatrix}\,.
\end{equation}
Using these relations, one can verify both that $\omega^{(k)}\cdot \omega^{\vee (l)}=A^{-1}_{kl}$, and also relation~(\ref{magicrel}). This is codified in the following matrix
\begin{equation}
\chi_{z*}(z)=e^{i 2\pi A_{kl}^{-1}}=
\left(
\begin{array}{ccc}
 1 & 1 & 1 \\
 1 & e^{-\frac{2 i \pi }{3}} &
   e^{\frac{2 i \pi }{3}} \\
 1 & e^{\frac{2 i \pi }{3}} &
   e^{-\frac{2 i \pi }{3}} \\
\end{array}
\right)\,,
\qquad\qquad
A_{kl}^{-1}= \frac 13
\begin{pmatrix}
0 & 0 & 0\\
0 & 2 & 1\\
0 & 1 & 2
\end{pmatrix}\,,
\end{equation}
where the extra row and columns of zeroes added to $A_{kl}^{-1}$ above with respect to the $SU(3)$ Cartan matrix represent the identity element and the identity representation of the center.

\subsection{Complementarity diagram in the non invariant space: a Maxwell analogue}
\label{com2}

In section \ref{com1} we introduced the relevant algebras for the non-Abelian theory, organized in the complementarity diagram (\ref{cus}). The relative entropies satisfy the certainty relation (\ref{cerinv}). As described in the preceding sections we can compute these relative entropies in a non-invariant Hilbert space. Since we do not have explicit gauge invariant smeared versions of line operators, we can use equation~(\ref{pok}) to relate the gauge invariant theory to the non-gauge invariant one, suggesting to bound the entropic order parameters using subalgebras of non-gauge invariant Wilson and 't Hooft loops belonging to the Cartan algebra, as described earlier.

We briefly summarize the situation in this setup with an analog in a theory of $l$ independent Maxwell fields, where we now impose the (Abelian) gauge invariant condition. In the non-Abelian theory, this gauge invariance will be imposed at the level of the expectation values.   
 Although the following is not strictly necessary in order to accomplish our objective, we notice we can build a complementarity diagram in the theory of multiple Maxwell fields by just including the non-gauge invariant Wilson and 't Hooft loops associated with the weight and co-weight lattices respectively:
\bea\label{wecowe}
{\cal A}_{W_\omega }(R) & \overset{E_W}{\longrightarrow} & {\cal A}_{W_\alpha }(R)\nonumber \\
\updownarrow\prime\! &  & \:\updownarrow\prime\\
\mathcal{A}_{W T_{\alpha^{\vee}}}(R') & \overset{E_T}{\longleftarrow} & {\cal A}_{W T_{\omega^{\vee}}}(R')\;.\nonumber 
\eea
This complementarity diagram has a similar flavor to the Maxwell-$Z_N$ case. 
In the upper left corner, we have the algebra with all WL in the full weight lattice and no TL.  In the lower right corner, we have the $R'$ local algebra plus the full weight lattice of TL and all WL. The subalgebras are defined by acting with the conditional expectations, which are defined by the chosen WL and TL in the Cartan subalgebra. This diagram then explicitly depends on the chosen group of non-local operators.  
 The conditional expectation kills all those operators which do not belong to the root lattice. 
  The index of this diagram is again $|Z|$, the dimension of the center because this is the number of independent elements in the weight lattice that are not in the root lattice. Bounds on the relative entropies associated with this diagram (in the theory of Maxwell fields) can be obtained by computing the relative entropies in the subalgebras of the relevant non-local operators. In turn, these will match the bounds on the non-Abelian theory in the weak coupling limit.   

In what follows, the objective will be to compute bounds on both entropic order and disorder parameters independently and verify the certainty relations. The bounds on the last diagram, obtained by using non-gauge invariant operators, translate to bounds on the real order parameters for the gauge theory.

\subsection{Bounds for WL order parameters}

Now we compute lower bounds on the order parameters in the weak coupling limit. To the lowest order in $g$, the expectation values arise by a theory of independent Maxwell fields with Lagrangian  
\be
L=-\frac{1}{4}\sum_a F_{\mu\nu}^a F^{\mu\nu}_a\,, \hspace{.7cm }F_{\mu\nu}^a= \partial_\mu A^a_\nu-\partial_\nu A^a_\mu \;.\label{ff}
\ee
Hence we will be able to use the same optimization described in section \ref{entropic}. 

Let us start with the WL order parameter, and simple groups, and then proceed to the general case.

\subsubsection{SU(2)}

The WL are labeled by the weight lattice. The conditional expectation takes us to the root lattice. For SU(2) the weights $\omega$ run over the half integers and the roots $\alpha$ over the integers. As Cartan algebra, we choose the third direction ($\sigma_3/2$) as usual. For a generic weight $\omega$, the Wilson loop expectation value is
\be
\langle W_{\omega} \rangle = \langle e^{i g \omega \int dx^3 J_i B_i}\rangle =e^{- \omega^2 \frac{g^2}{2}\langle \Phi_B^2\rangle }\;.
\ee
Since the weight lattice WL form a simple abelian algebra
\be 
W_{\omega}W_{\omega '}=W_{\omega+\omega '}\;,
\ee
we can diagonalize this algebra using a conventional Fourier transform. In particular the projectors are
\be 
p_x=\frac{1}{2\pi}\sum_{n\in\mathbb{Z}}e^{i n x}\, (W_{1/2})^n\;.
\ee
Defining again $c^2=\frac{g^2}{2}\langle \Phi_B^2\rangle$, and considering the weak coupling limit $c\ll 1$, the expectation values in the vacuum $\omega$ are\footnote{We apologize for the overlap in notation, where $\omega$ means the vacuum in this case but above it was referring to weights in the weights lattice. Since the meaning of the symbol is always transparent, we have decided not to modified the otherwise widely known and accepted conventions for these two notions.}
\begin{equation}
p_x^{\omega}=\,\frac{1}{2\pi}\,\sum_{n\in\mathbb{Z}}e^{i n x}\,\langle W_{1/2} \rangle^{n^2}\sim \frac{e^{-\frac{x^2}{c^2}}}{ \sqrt{\pi}c}\;,
\end{equation}
while for the vacuum composed with the conditional expectation
\begin{equation}\label{eqg1}
p_x^E=\,\frac{1}{2\pi}\,\sum_{n\in\mathbb{Z}}e^{i 2 n x}\,\langle W_{1/2} \rangle^{4n^2}\sim \frac{1}{2}\left( p_x^{\omega} + p_{x-\pi}^{\omega} + p_{x+\pi}^{\omega} \right)\;.
\end{equation}
The order parameter is defined as
\be
S_W=\int_{-\pi}^{\pi} p_x^\omega \ln\left( \frac{ p_x^\omega}{ p_x^E}\right) \sim \ln(2) -\frac{2 c }{\pi ^{3/2}}e^{-\frac{\pi ^2}{4 c^2}}\,. 
\ee
Notice that exact calculation of the expectation value for the projectors in terms of $\langle W_{1/2}\rangle$ was possible. However, the computation so far mirrors that of the Maxwell case. Thus, following the discussion below eq. \eqref{STMax222}, one can see that getting the position of the Gaussian peaks is enough to compute the scaling of the order parameter in this approximation.
This can be checked numerically in a straightforward manner.

\subsubsection{SU(3)}

\begin{figure}[t]
\includegraphics[width=.6\linewidth]{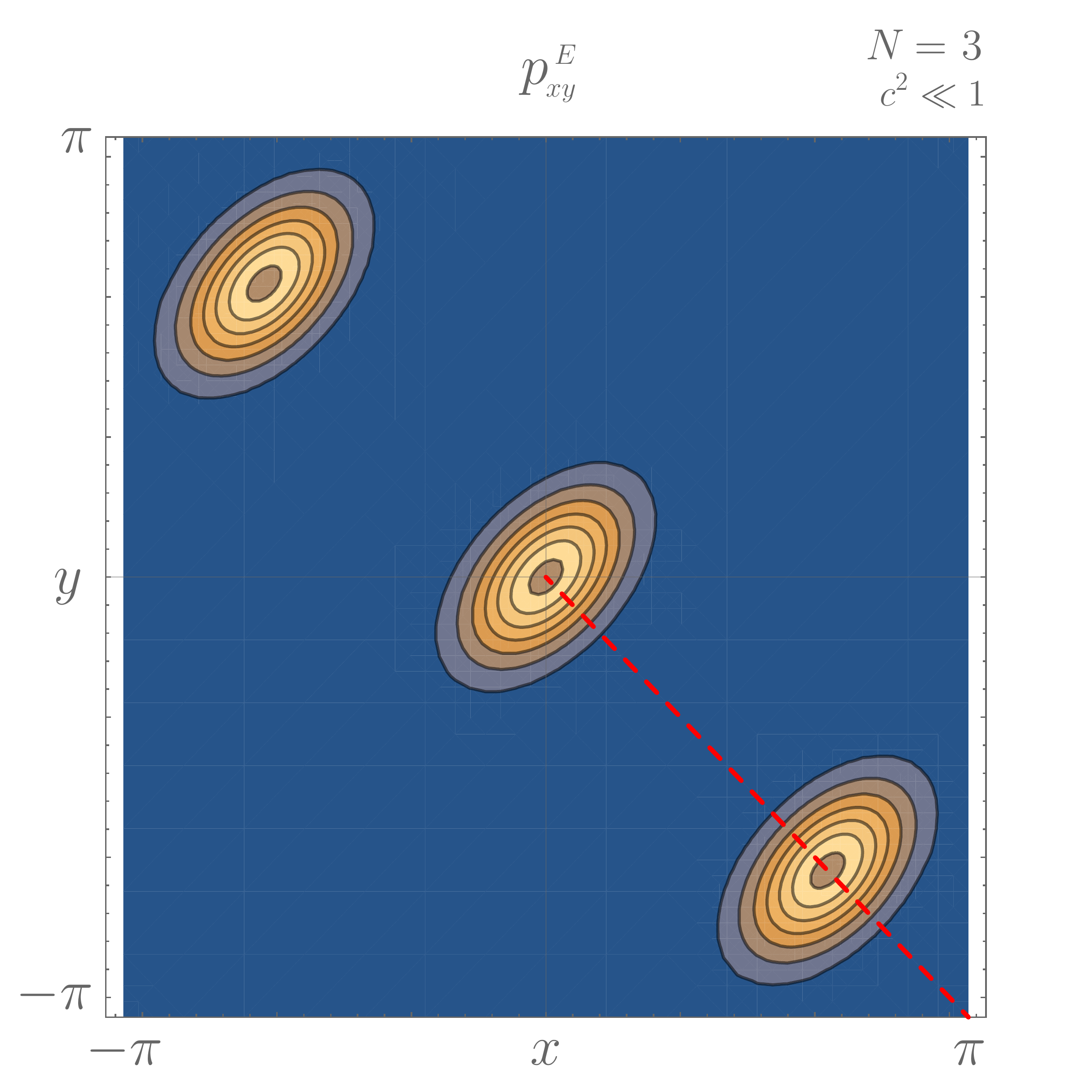} 
\centering
\caption{A contour plot of $p_{xy}^{E}$ is presented. Notice that the structure of Gaussian peaks seen in all previous scenarios is also present but in a diagonal direction combining $x$ and $y$. If we take the dashed line as an axis one gets Fig. \ref{Fig:Gauss}. }
\label{Fig:e3f}
\end{figure}

In SU(3) the weight and root lattices are 2 dimensional. The fundamental weights are
\be
\omega^{(1)}=\frac{1}{6}(3,\sqrt{3})\,,\,\,\,\,\,\,\,\,\,\,\,\,\,\,\,\,\omega^{(2)}=\frac{1}{6}(0,2\sqrt{3})\;,
\ee
with norms equal to $1/3$. The fundamental roots are 
\be
\alpha^{(1)}=(1,0)=2\omega^{(1)}-\omega^{(2)}\,,\,\,\,\,\,\,\,\,\,\,\,\,\,\,\,\,\alpha^{(2)}=\frac{1}{2}(-1,\sqrt{3})=2\omega^{(2)}-\omega^{(1)}\;.
\ee
The WL, taking values in the weight lattice, are labeled by a pair of integers, i.e.
\be 
\omega=n\,\omega^{(1)}+m\,\omega^{(2)}\,\,\,\,\Longrightarrow\,\,\,\,\, \vert\omega\vert^2=\frac{n^2+ nm + m^2}{3}\,.
\ee
The algebra is abelian once again, so
it can be diagonalized by Fourier transforming in both directions. The projectors are
\be 
p_{xy}=\,\frac{1}{4\pi^2}\,\sum_{n,m\in\mathbb{Z}}e^{i n x}e^{i m y}\, W_{\omega_{nm}}\;.
\ee
We choose the magnetic fluxes in both Cartan direction to be equal. The expectation values of the Wilson loops are, with $c^2=\frac{g^2}{2}\langle \Phi_B^2\rangle$, 
\begin{equation}
\langle W_{\omega_{nm}} \rangle 
=e^{-c^2 \vert\omega\vert^2}\;.
\end{equation}
We can now compute the expectation values of the projectors in both states. In the vacuum we have
\begin{equation}
p_{xy}^{\omega}=\,\frac{1}{4\pi^2}\,\sum_{n ,m\in\mathbb{Z}} e^{i n x}e^{i m y}e^{-c^2 \vert\omega\vert^2}\sim \frac{ \sqrt{3} }{2\pi  c^2} e^{-\frac{ x^2-x
   y+y^2}{c^2}}\;.
\end{equation}
In the vacuum composed with the conditional expectation, an analogous reasoning in the root lattice leads to
\begin{align}\label{eqg2}
p_{xy}^E&=\,\frac{1}{4\pi^2}\,\sum_{n ,m\in\mathbb{Z}} e^{i (2n-m) x}e^{i (2m-n) y}e^{-c^2 (n^2- nm + m^2)}\nonumber\\
&\sim\frac{1}{3}\left( p_{xy}^{\omega}+ p_{(x+\frac{2\pi}{3}),(y-\frac{2\pi}{3})}^{\omega}+ p_{(x-\frac{2\pi}{3}),(y+\frac{2\pi}{3})}^{\omega}\right) \;.
\end{align}
We recover again the Gaussian structure but this time in 2 dimensions. In Fig~(\ref{Fig:e3f}), we plot \eqref{eqg2} to provide an intuitive picture of its form.
The order parameter is
\be\label{r3}
S_W=\int_{-\pi}^{\pi} p_{xy}^{\omega} \ln\left( \frac{p_{xy}^{\omega}}{p_{xy}^{E}}\right)\sim \ln(3) -\frac{c }{\sqrt{3}\pi ^{3/2}}e^{-\frac{\pi ^2}{3 c^2}}\;.
\ee
Notice that despite being a 2-dimensional problem in this case the structure of displaced Gaussians appears again, albeit tilted in the $\{x,y\}$ plane. This suggests the existence of an underlying structure for general gauge groups, which we present below.

\subsubsection{General gauge groups}
\label{uN}

What we learned in the calculations for the simplest groups allows us to understand the case of a general gauge group $G$. For a general group with rank $l$, there are $l$ fundamental weights $\omega^{(i)}$. A generic weight is then parametrized by $l$ integer numbers $n_1,n_2,\cdots,n_l$ as
\be 
\omega=n_1\omega^{(1)}+n_2\omega^{(2)}+\cdots +n_l\omega^{(l)}\;.
\ee
Luckily, the weight lattice WL algebra still forms an abelian algebra
\be 
W_{\omega}W_{\omega '}=W_{\omega+\omega '}\;.
\ee
Therefore it can be diagonalized by doing $l$ Fourier transforms along the different directions set by the different fundamental weights
\be 
p_{x_1x_2\cdots x_l}\equiv \frac{1}{(2\pi)^l}\sum_{n_1,n_2,\cdots, n_l\in\mathbb{Z}}e^{i \sum_i n_i x_i}\, W_{\omega}\;.
\ee
Now the first step is to understand the distribution $p_{x_1x_2\cdots x_l}^\omega$ produced by the vacuum. The expectation value for the WL is given by
\be
\langle W_{\omega}\rangle = e^{-c^2\, \vert\omega\vert^2}=e^{-\frac{c^2}{2}\, \sum_{ij}\,n_i \,M_{ij} \,n_j}\,, 
\ee
where we have defined the matrix $M_{ij}=2\,\omega^{(i)}\cdot\omega^{(j)}$. This implies
\be
p_{x_1,\cdots ,x_l}^{\omega} =\frac{1}{(2\pi)^l} \sum_{n_1,n_2,\cdots, n_l\in\mathbb{Z}}e^{i \sum_i n_i x_i}\, e^{-\frac{c^2}{2}\, \sum_{ij}\,n_i \,M_{ij} \,n_j}\,.
\ee
In the small coupling limit, this has a Gaussian form concentrated around the origin. Near the origin we could take a continuum limit, transforming the sum into an integral. We obtain
\be
p_{x_1,\cdots ,x_l}^{\omega}\simeq\frac{1}{(2\pi)^l} \int \prod_i dn_i \,e^{i \sum_i n_i x_i}\, e^{-\frac{c^2}{2}\, \sum_{ij}\,n_i \,M_{ij} \,n_j}= \frac{1}{\sqrt{(2\pi)^l \,\textrm{det} M}}e^{-\frac{1}{2c^2}\sum_{ij}\,x_i \,M_{ij}^{-1} \,x_j}\,.
\ee   
Let us now consider the vacuum composed with the conditional expectation. This just means the projectors are computed by summing only over the root lattice. The relation between fundamental roots and weights was
\be 
\alpha^{(i)}=\sum_j A_{ij}\,\omega^{(j)}\;.
\ee
A generic root is then
\be 
\sum_i q_i\alpha^{(i)}=\sum_i q_i\sum_j A_{ij}\,\omega^{(j)}=\sum_j (\sum_i q_i A_{ij}) \,\omega^{(j)}\;.
\ee
This means we sum over weights satisfying $n_j=\sum_i q_i A_{ij}$ so that
\be 
p_{x_1x_2\cdots x_l}^E\equiv \frac{1}{(2\pi)^l}\sum_{q_1,q_2,\cdots, q_l\in\mathbb{Z}}e^{i \sum_j (\sum_i q_i A_{ij}) x_j}\, e^{-\frac{c^2}{2}\, \sum_{jl}\,(\sum_i q_i A_{ij}) \,M_{jl} \,(\sum_k q_k A_{kl})}\;.
\ee
Again, near the origin we can approximate the sum by an integral. Within the integral we can change variables from the $q_i$ to the $n_i=\sum_j q_j A_{ij}$. The Jacobian of this transformation is $\textrm{det} A=\vert Z\vert$ and we arrive at
\be 
p_{x_1x_2\cdots x_l}^E\simeq \frac{1}{\vert Z\vert}\,p_{x_1,\cdots ,x_l}^{\omega}\;.
\ee
The prefactor should be compared with the prefactor in eq.~(\ref{eqg1}) for $SU(2)$ and eq.~(\ref{eqg2}) in $SU(3)$. When inserted in the relative entropy, the $1/\vert Z\vert$ will give rise to the topological leading $\log \vert Z\vert $ contribution.

We now need to obtain the subleading term. To such end, we need to understand the symmetries of each of the distributions, and concretely the symmetries of the second one. As it is well known in condensed matter physics, whenever we have a function that is symmetric under translations in a given lattice, the Fourier transform contains momentum modes that belong to the dual reciprocal lattice. Going in the reverse direction, if we Fourier transform a function that runs over a given lattice, the transformed function will be described solely by the modes belonging to the first Brillouin zone of the reciprocal lattice. The dual function will then be symmetric under translation over the reciprocal lattice.

Let us remind the definition of the reciprocal lattice. Given a lattice generated by $\omega^{(i)}$, the reciprocal lattice is spanned by vectors $e^{(i)}$ satisfying
\be 
\omega^{(i)}\cdot e^{(i)}=2\pi\delta_{ij}\;.
\ee
If $\omega^{(i)}$ are the generators of the weight lattice, the previous equation means that $e^{(i)}=2\pi \alpha_{(i)}^{\vee}$ generate the reciprocal lattice. In other words, the reciprocal lattice to the weight lattice is $2\pi$ times the co-root lattice. Similarly, the reciprocal lattice to the root lattice is $2\pi$ times the co-weight lattice.

Now, since the first distribution $p_{x}^{\omega}$ is a sum over the weight lattice, it is invariant under translations over $2\pi$ times the co-root lattice. Since we can write the exponents of the Fourier transform as a inner product
\be 
\sum_i n_i x_i = \omega\cdot\alpha^{\vee}\;,
\qquad\qquad
\omega\equiv\sum_i n_i\, \omega^{(i)}\;,
\qquad\qquad
\alpha^{\vee}\equiv \sum_i x_i \,\alpha^{\vee}_{(i)}\;,
\ee
we see that, when translated to the $x$ dependence, this symmetry means, as usual, that the distribution is invariant under $2\pi$ translations of the $x$ coordinates. 

On the other hand, the second distribution only contains a sum over the root lattice. This means it is invariant under $2\pi$ times the co-weight lattice. This is a bigger group of symmetries than the previous one since the co-weight lattice contains the co-root lattice. This means that, apart from the peak at the origin, in the first Brillouin zone we have $|Z|-1$ further peaks which are equal to the peak at the origin. The position of these peaks will be
\be 
\sum_i\Delta x_i \,\alpha^{\vee}_{(i)}=2\pi \omega^{\vee}\qquad\Longrightarrow\qquad \Delta x_i =2\pi \omega^{\vee}\cdot\omega^{(i)}\;.
\ee
The probability distribution of the vacuum composed with the conditional expectation has the same form at these positions that it has at the origin due to the enlarged symmetry and we then find
\be 
p_{\Delta x}^E\sim \frac{1}{\vert Z\vert} e^{-\frac{(2\pi)2}{2c^2} \,\omega^{\vee}\cdot(\sum_{ij}\omega^{(i)}M_{ij}^{-1}\omega^{(j)})\cdot \omega^{\vee}}\;.
\ee
Notice that 
\be 
M_{ij}=2\,\omega^{(i)}\cdot\omega^{(j)}=\vert\alpha^{(j)}\vert^2 A^{-1}_{ij}\Longrightarrow M_{ij}^{-1}=\frac{1}{\vert\alpha^{(j)}\vert^2}A_{ij}\;.
\ee
Then
\be 
\sum_{ij}\omega^{(i)}M_{ij}^{-1}\omega^{(j)}=\frac{1}{2}\sum_j \alpha^{\vee}_j\omega^{(j)}=\frac{1}{2}\,\mathds{1}_{l\times l}\;.
\ee
We conclude that the second probability distribution at the symmetry translated peaks is
\be 
p_{\Delta x}^E\sim \frac{1}{\vert Z\vert} e^{-\frac{(2\pi)^2}{4c^2} \,\omega^{\vee}\cdot \omega^{\vee}}\;.
\ee
The result for the relative entropy then follows in the same way as in the calculations above. The leading term is $\log |Z|$ and it comes from the comparison between $p_x^{\omega}$ and the piece $|Z|^{-1} p_x^{\omega}$ contained in $p_x^{E}$. The leading exponential term in the correction comes from the center vector $\omega^{\vee}_*$ which is nearest in norm to the identity, $\vert\omega^{\vee}_*\vert^2=\textrm{min}_z \vert\omega^{\vee}_z\vert^2 $. It corresponds to evaluate the vacuum probability distribution halfway between the identity and $\omega^{\vee}_*$:
\be
S_W= \log |Z|- \alpha\, e^{- \frac{(2\pi)^2}{16 c^2}\vert\omega^{\vee}_*\vert^2 }\,   \label{nuevaw}
\ee
where $\alpha$ is some coefficient at most polynomial in the coupling constant. 

As an example, let us evaluate $e^*\equiv2\pi\omega^{\vee}_*$ for $SU(3)$ and check the result of the previous section.  The center $\mathbb{Z}_3$, and generically the center of any group $G$, can be generated by only exponentiating linear combinations of its Cartan subalgebra, i.e. its diagonal elements. In the Gell-Mann basis, these are
\begin{equation}
\lambda_3=
\begin{pmatrix}
1 & 0 & 0\\
0 & -1 & 0\\
0 & 0 & 0
\end{pmatrix}\,,
\qquad\qquad
\lambda_8=
\frac{1}{\sqrt{3}}\begin{pmatrix}
1 & 0 & 0\\
0 & 1 & 0\\
0 & 0 & -2
\end{pmatrix}\,.
\end{equation}
In $SU(3)$ we have two non-trivial center elements, i.e. $e^{\pm i\frac{2\pi}{3}}$  where the $3\times3$ identity is left implicit. We thus call $e_\pm^a$ the vectors in the Lie algebra that reproduce these center elements, i.e.
\begin{equation}
e^{\frac i2\left( e_\pm^3 \lambda_3+e_\pm^8 \lambda_8\right)} = e^{\pm i\frac{2\pi}{3}} 
\qquad\Rightarrow\qquad
e_\pm^a=(e_\pm^3,e_\pm^8)=\mp \frac{2\pi}{\sqrt{3}}(\sqrt 3,1)
\,.
\end{equation}
We see that we have two charges with the same modulus $(e^*)^2=\frac{4}{3}(2\pi)^2$. Upon replacing in (\ref{nuevaw}) gives the result (\ref{r3}).

For general $SU(N)$ the computation is straightforward to generalize by induction. $SU(N)$ has $N-1$ elements in the Cartan subalgebra, which can in turn be chosen to be those of $SU(N-1)$ with a extra rows and columns of zeroes to match dimensions, plus an extra diagonal element of the form 
\begin{equation}
\frac{1}{\sqrt{N(N-1)}}\begin{pmatrix}
1 & \cdots & 0 &0\\
\vdots & \ddots & 0& 0\\
0 & 0 & 1& 0\\
0 & 0 & 0 &  -N(N-1)
\end{pmatrix}\,.
\end{equation}
Notice that $\lambda_8$ in eq. \eqref{GellMann} is an example of this recursive construction. The smallest charge $e^*$ will always be associated to the non-trivial center element closest in argument to the identity\footnote{One can always choose an element $\tilde \lambda$ in the basis of the Cartan subalgebra such that it can generate the group center on its own. This is, the group center is generated by $e^{2\pi i \sqrt{\frac{2(N-1)}{N}} j \tilde \lambda}$ with $\lfloor \frac{N-2}{2} \rfloor\leq j \leq \lfloor \frac{N}{2}\rfloor$, $j\in\mathbb{Z}$. In this basis, it is immediate to see that the $j$-th center element distance to the identity is $\frac{2(N-1)}{N} (2\pi)^2 j^2$, of which its minimal non-trivial value is precisely $(e^*)^2$, at $j=\pm1$. For example, in the standard notation for the $SU(4)$ algebra generators, $\tilde \lambda=-\sqrt{\frac{2}{3}} \left(\lambda _3+\frac{\lambda   _8}{\sqrt{3}}+\frac{\lambda  _{15}}{\sqrt{6}}\right)$. }, i.e. $e^{i\frac{2\pi}{N}}$. Using these matrices to match this center element leads to
\begin{equation}
e^*= -2\pi\left\{1,\frac{1}{\sqrt{3}},\dots,\frac{1}{\sqrt{N(N-1)}}\right\}
\quad\Rightarrow\quad
(e^*)^2 = \frac{2(N-1)}{N} (2\pi)^2 \,.
\end{equation}
 Therefore for $SU(N)$ we have
\begin{align}\label{guess}
S_{W}
\sim  \ln N - 
\alpha\, e^{-\frac{N-1}{N}\frac{\pi ^2}{2 c^2}}\;.
\end{align}
Despite its similarities with the Maxwell scenario, notice that weight and co-weight lattice structure makes the $N\to\infty$ limit finite, in strong contrast with the Maxwell $Z_n$ scenario.

\subsection{Lower bounds for the TL order parameter}
\label{entd}

We move on to the entropic disorder parameter. Following the complementarity diagram \eqref{cus}, this is defined by the following relative entropy
\be 
S_{ {\cal A}_{WT}(R)}(\omega,\omega\circ E_{T})\;.
\ee
This disorder parameter quantifies the amount of generalized symmetry generated by the 't Hooft loops. This relative entropy is computed on the maximal algebra (the additive algebra of the ring plus the WL plus the TL). It quantifies the distance between the vacuum state and the vacuum state composed with the dual conditional expectation $E_T$. The conditional expectation projects the maximal algebra into the additive algebra plus the WL, i.e. killing the TL.

We again use monotonicity of relative entropy and compute a lower bound
\be 
S_{{\cal A}_{WT}(R)}(\omega,\omega\circ E_{T})\geq S_{{\cal A}_{lower}(R)}(\omega,\omega\circ E_{T})\;,
\ee
where ${\cal A}_{lower}(R)\subset{\cal A}_{WT}(R)$ and more concretely we will choose the algebra generated by a single 't Hofft loop. 
In other words, an operator $\mathcal{O}\in {\cal A}_{lower}(R)$ can be written as
\be 
\mathcal{O}=\sum\limits_{n} a_{n} T^n\;,
\ee
where $T^n$ is the $n$'th power of a TL corresponding to a non-trivial element $z$ of the center of the gauge group, which was defined in~(\ref{t22}) and~(\ref{t222}). We have included all powers because we remind that in the process of smearing the TL, we lost the product algebra properties of the TL, so to close an algebra we need in principle all integer powers. However, notice that even if the TL powers do not satisfy the group operations, the equivalence class to which they belong reproduces the operations of the group, as can be verified by computing the commutator with dual WL. In other words, $T^Z$, although not the identity operator exactly, is in the identity class and commutes with all WL outside. That is the reason we only need the algebra generated by $T$. This will test a non-trivial cycle inside the center of the group. 

In practice though, as was also the case for the Maxwell scenario, only the TL corresponding to the minimum charge in the center contributes to the weak coupling regime. Basically, in such a regime all TL expectation values are exponentially close to zero and we need only the leading exponential term.

\subsubsection{SU(2)}
\label{u2d}

Let's start with $SU(2)$. In this case, there is only one non-trivial TL, call it $T$. Any operator in the algebra generated by $T$ can be written as
\be 
\mathcal{O}=\sum\limits_n a_{n} T^n\;.
\ee
The conditional expectation kills the non-local TL. These are the TL that do not commute with some WL in the complementary region. In the $SU(2)$ case, this means it kills all the odd powers of $T$.
Defining as usual $\tilde c^2=\frac 12 \left(\frac{2\pi}{g}\right)^2 \langle\Phi_E^2\rangle$, the expectation value of the TL and its powers is
\begin{equation}
T = e^{\frac ig \int dx^3  J_i^a \tilde A^a_i}
\quad \Rightarrow \quad 
\langle T^m \rangle = e^{-\frac 12 m^2 \left(\frac{2\pi}{g}\right)^2 \langle\Phi_E^2\rangle}=e^{-m^2 \tilde c^2}\,.
\end{equation}
This abelian algebra is given by integer powers of the lowest charge, like in the Maxwell field case. The projectors are
\be 
\tilde p_x=\frac{1}{2\pi}\sum_n e^{inx} T^n\;.
\ee 
In the small coupling limit the expectation values of the proyectors in the vacuum are then
\begin{equation}
\tilde p_x^{\omega}=\frac{1}{2\pi}\sum_{m\in\mathbb{Z}}e^{inx}\langle T^m \rangle 
\sim \frac{1}{2\pi}+\frac{1}{\pi} e^{-\tilde{c}^2} \cos (x)\;,
\end{equation}
while for the vacuum composed with the conditional expectation
\begin{equation}
\tilde p_x^{E}=\frac{1}{2\pi}\sum_{m\in\mathbb{Z}}e^{i2nx}\langle T^{2m} \rangle
\sim \frac{1}{2\pi}+\frac{1}{\pi} e^{-4 \tilde{c}^2} \cos (2x )\;.
\end{equation}
We can now compute the lower bound by computing the classical relative entropy between these probability distributions
\begin{equation}
S_{T}=\int_{-\pi}^\pi dx\, \tilde p_x^{\omega}\ln \left(\frac{\tilde p_x^{\omega}}{\tilde p_x^{E}}\right) \sim e^{-2 \tilde{c}^2}=e^{-\left(\frac{2 \pi }{g}\right)^2 \langle\Phi_E^2\rangle}\;.
\end{equation}
In the $\tilde c^2\gg1 $ limit, as expected, the only relevant contribution comes from the smallest charge TL VEV. The lower bound is given simply by such VEV squared because the relative entropy cannot be sensitive to a change of sign in the operator.

\subsubsection{General gauge groups}
\label{uNd}

We have gone through the complete calculation for $SU(2)$ but the result follows directly from perturbation theory, taking into account that the only relevant contribution to the leading exponential order comes from the TL with the smallest charge. In regards to the relative entropy calculation, we can as well think we have an algebra with two elements, the identity, and $T$, in the limit $\langle T \rangle \ll 1 $. We then evaluate the relative entropy between two states giving $\langle T \rangle$ and zero expectation value to the TL respectively. This relative entropy goes as $\sim \langle T \rangle^2$. 

The result for general gauge groups is then easily obtained. We have to compute the expectation value of the TL with the highest VEV, and the relative entropy is determined by the square of this VEV, to leading exponential order.  

Using~(\ref{t222}) for the TL
\be
T_{\omega^{\vee}}=e^{i \frac{2\pi}{g}\,\int d^3x\,  j_i \,\omega^{\vee}_a \, E_i^a}=e^{i \frac{2\pi}{g}\,\int d^3x\,  J_i \,\omega^{\vee}_a \, \tilde{A}_i^a}\,,
\ee
the leading contribution corresponds to the coweight $\omega^{\vee}_*$ with the smallest norm. This has expectation value
\be
\langle T_{\omega^{\vee}}\rangle= e^{- \frac{(2\pi)^2}{2 g^2} \vert\omega^{\vee}_*\vert^2 \langle\Phi_E^2\rangle}\;.
\ee
Therefore the general result is 
\be
S_T\sim e^{-\frac{(2\pi)^2}{ g^2} \vert\omega^{\vee}_*\vert^2 \langle\Phi_E^2\rangle }=e^{- 2\vert\omega^{\vee}_*\vert^2 \tilde{c}^2}\,.
\ee 
In particular, for $SU(N)$ theories we have in the $\tilde c^2\gg1 $ limit
\begin{equation}
\langle T_{\omega^{\vee}}\rangle= e^{- \frac{N-1}{N} \left(\frac{2\pi}{g}\right)^2 \langle\Phi_E^2\rangle}\,,\hspace{.7cm}
S_{T} \sim e^{- 2\frac{N-1}{N} \left(\frac{2\pi}{g}\right)^2 \langle\Phi_E^2\rangle}\,.
\end{equation}
Notice again that in the $N\to\infty$ limit the bound exponent remains finite, in contrast with the Maxwell $\mathbb{Z}_n$ scenario.

\subsection{Certainty relation and summary}
\label{cert}

The complementarity diagram for gauge theories that we have studied is 
\bea\label{cus11}
{\cal A}_{W}(R) & \overset{E_W}{\longrightarrow} & {\cal A}_{\text{add}}(R)\nonumber \\
\updownarrow\prime\! &  & \:\updownarrow\prime\\
\mathcal{A}_{W}(R') & \overset{E_T}{\longleftarrow} & {\cal A}_{WT}(R')\;.\nonumber 
\eea
The index associated with the dual conditional expectations $E_{W}$ and $E_{T}$ is $|Z|$, which is just the number of independent WL and TL killed in $R$ and $R'$ respectively. The associated certainty relation is
\be 
S_{{\cal A}_{W}(R)}(\omega,\omega\circ E_{W})+S_{ {\cal A}_{WT}(R')}(\omega,\omega\circ E_{T})=\log |Z|\;.
\ee
As explained above, this can be used to provide upper bounds to the relative entropies employing the lower bounds. In this case, the inequalities take the following form
\bea\label{boundsud2}
S_{W(R)}(\omega,\omega\circ E_W)&\leq & S_{{\cal A}_{W}(R)}(\omega,\omega\circ E_W)\leq \log |Z| -S_{T(R')}(\omega,\omega\circ E_T)\,, \\
S_{T(R')}(\omega,\omega\circ E_T)&\leq & S_{{\cal A}_{WT}(R')}(\omega,\omega\circ E_T)\leq \log |Z| -S_{W(R)}(\omega,\omega\circ E_W)\;.
\eea
Using the results of the previous sections, we can now verify these inequalities.

We have shown that 
\be
e^{- \frac{(e^*)^2}{ g^2} \langle \Phi_E^2\rangle(R')}\le \log |Z|- S_{{\cal A}_{W}(R)}(\omega,\omega\circ E_{W})=S_{ {\cal A}_{WT}(R')}(\omega,\omega\circ E_{T})\le \alpha\, e^{- \frac{(e^*)^2}{8 g^2 \langle \Phi_B^2(R)\rangle} }\,,\label{diff}
\ee
 with $\alpha$ some constant with at most power law dependence  on the coupling constant, and $(e^*)^2=(2\pi\omega^{\vee}_*)^2$. 
These inequalities are consistent because we have the inequality for the non-commuting Gaussian electric and magnetic modes
\be
\langle\Phi_E^2\rangle(R')\langle\Phi_B^2\rangle(R)\ge\frac{1}{4}\,.
\ee
Because of electromagnetic duality for the Maxwell field, the optimal electric and magnetic fluxes are the same functions of the region, $\langle\Phi_E^2\rangle(R)=\langle\Phi_B^2\rangle(R)$. In (\ref{diff}) they are evaluated for complementary regions. 
When inequality (\ref{diff}) saturates the upper and lower bounds in (\ref{diff}) differ only by a factor $2$ in the exponent. In the vacuum state, saturation occurs in the limit of very wide or thin loops.
 
The inequalities (\ref{diff}) contain the main result of this paper. They imply the order parameter for the WL is exponentially near saturation $\log|Z|$ and the one for the TL is exponentially small in the inverse of the coupling constant squared. The coefficient on the exponent depends on the group through the minimal monopole charge squared $(e^*)^2=(2\pi\omega^{\vee}_*)^2$, and the geometric dependence is constrained by the optimal fluxes which are computable exclusively in terms of the Maxwell theory. We have described these geometric functions in section \ref{eno}. Let us call $\tilde{l}$ and $\tilde{r}$ to the length and width of the ring where the 't Hooft loop is included. That is, these are taken here as the dimensions of $R'$ rather than of $R$, and are related to the dimensions of the Wilson loop ring $R$ by the relation of cross ratios $\eta'=1-\eta$, see section \ref{com}. For the case of  a thin TL ($\tilde{l}/\tilde{r}\gg 1$)  loop  we have
\be
e^{- \frac{\pi}{4}\frac{(2\pi)^2}{ g^2} (\omega^{\vee}_*)^2\frac{\tilde{l}}{\tilde{r}}}\le S_{ {\cal A}_{WT}(R')}(\omega,\omega\circ E_{T})\le \alpha\, e^{- \frac{\pi}{8}\frac{(2\pi)^2}{ g^2} (\omega^{\vee}_*)^2 \frac{\tilde{l}}{\tilde{r}}}\,.\label{diffaa}
\ee
This gives a perimeter law for thin TL and fixes the coefficient of the perimeter within a factor $2$. For wide loops ($\tilde{r}/\tilde{l}\gg 1$) we have instead
\be
e^{- \frac{\sqrt{2}}{\pi}\frac{(2\pi)^2}{ g^2} (\omega^{\vee}_*)^2 \sqrt{\frac{\tilde{l}}{\tilde{r}}}}\le S_{ {\cal A}_{WT}(R')}(\omega,\omega\circ E_{T})\le \alpha\, e^{- \frac{1}{\sqrt{2}\pi}\frac{(2\pi)^2}{ g^2} (\omega^{\vee}_*)^2 \sqrt{\frac{\tilde{l}}{\tilde{r}}}}\,.\label{diffma}
\ee
The WL order parameter of course remains exponentially close to the saturation constant $\log|Z|$ in all the range. 

We make some final comments. The first concerns results for different order parameters that can be constructed either by eliminating WL or TL. Notice that according to our calculation of the bounds, eq. (\ref{diff}) does not change if we simultaneously replace
\be
 S_{{\cal A}_{W}(R)}(\omega,\omega\circ E_{W})\rightarrow S_{{\cal A}_{WT}(R)}(\omega,\omega\circ E_{W})\,,\hspace{.6cm} S_{{\cal A}_{WT}(R')}(\omega,\omega\circ E_{T})\rightarrow S_{{\cal A}_{T}(R')}(\omega,\omega\circ E_{T})\,.\label{piu}
\ee
The reason is that the main contribution to these relative entropies come from the difference between states due to some non-local operator being set to zero. In (\ref{piu}) the operators that are set to zero by the conditional expectation are unchanged. To be sensitive to differences between these order parameters, as well as to improve the bounds (\ref{diff}), it is necessary to include additive operators in the algebra leading to the upper and lower bounds. We will not pursue this calculation in this paper. 

The second comment is about the range of validity of the inequalities. We have assumed a fixed geometry and the limit of a small charge. This is why the WL is almost saturated. This means we have to keep $g^2 l/r\ll 1$, otherwise, our calculations of the WL will be invalidated. Perturbative corrections will also start to be relevant for $g^2 l/r\gtrsim 1$. In the same way, the complement of thin WL, e.q. (\ref{diffma}), is valid as far as the loop is not wide enough, $\sqrt{\tilde{l}/\tilde{r}}\gg g^2$. 
 Hence, we do not have access to the perimeter law for the WL (which must be valid for small enough $r/l$) while we have access to this law for the TL. 
The perimeter law for the WL will be subject to the RG flow and in the limit of line operators we should regain path-ordered loops. 
 
Finally, it is not difficult to see the changes to our results when we add matter fields to the Lagrangian, while remaining in the weak coupling regime, without symmetry breaking. The matter fields will modify the expectation values of non-local operators in a perturbative way, so they do not change the leading behavior in our expressions. However, electrically charged particles in a representation that is not generated by the adjoint representation will transform a non-local operator into a local one, since the corresponding Wilson line may be used to break the WL into local operators. At the same time, these electric charges convert the non-commuting TL into surface operators. An analogous reduction of the group of allowed non-local operators results from adding monopoles. See for example \cite{Casini:2020rgj}. The results of this paper apply equally well to these cases with the only difference that the relevant minimal charge $e^*$ and group of non-local operators (a subgroup of the center of the gauge group) have to be computed, after taking into account this reduction.

\section{Discussion}
\label{flow}
 
 Our principal results were already summarized in section \ref{cert}. Here we make two final remarks. The first one is to highlight a new idea about smeared non-local operators that has been suggested by the computation of the relative entropy order parameters. 
 As it is well known, in the physics literature Wilson and 't Hooft loops are defined as line operators by specifying a representation of $G$ and a representation of its dual respectively, see \cite{Kapustin_2006}. This is the same as specifying a dominant weight and a dominant co-weight. In the present article, due to the need of defining smeared loops, we have arrived at a new characterization in a non gauge invariant Hilbert space, based on the full weight and co-weight lattices. These loops are then not associated with representations of $G$ and its dual, but they are still classified by representations of the center of the group and its dual (the center of the group itself). At the gauge invariant level it is expected that non-local smeared operators could only be classified by their truly non-local classes $Z$ and $Z^*$.

Our second comment is about RG flows. Relative entropy order parameters give a new perspective on generalized symmetries and their dynamical breaking. An important objective is to understand the behavior of these quantities with the renormalization group or, in other words, with size. We comment on what the present calculation can teach us in this regard.

Suppose we are following the RG trajectory of a non-Abelian theory in the weak coupling regime, and that there is only one coupling $g$. We have the RG formula
\be
\mu\frac{d g}{d\mu}=\beta(g)\,.    
\ee 
The beta function can have any sign, but as we move along the RG trajectory this sign cannot change because at the changing point $\beta(g)=0$ and we hit a CFT fix point. 

In our formulas for the order parameters in the weak coupling regime, we should replace $g$ by $g(r)$, where $r$ is a typical size of the region that changes as we scale it in size, keeping its overall geometric form invariant. We have already observed that our formulas are valid provided  $g$ times any ratio of scales in the geometry is small enough. Otherwise, if $g(r)/g(l)$ differ significantly from $1$, for two different scales $r,l$ of the geometry into consideration, the perturbative corrections start to be important, since we cannot take a unique value of $g$ for the geometry.   

 The monotonicity of the coupling constant with scale translates to our order parameters in the weak coupling regime directly because they are monotonic functions of the coupling constant. 
The dual order parameters have opposite behaviour with scale, when one grows the other must decrease. 
For example, in the asymptotically free regime $g^2(r)= c/\log(r_0/r)$, $r\ll r_0$, and we get  
\be
 \log |Z|- S_{{\cal A}_{W}(R)}(\omega,\omega\circ E_{W})=S_{ {\cal A}_{WT}(R')}(\omega,\omega\circ E_{T})\sim  e^{- f \log(r_0/r)}=\left(\frac{r}{r_0}\right)^{f}\,.
\ee 
This is a power-law behavior with the scale that characterizes the asymptotically free regime, where the exponent $f$ is a function of the geometry of the ring (but not of its overall size) which we have constrained in equation (\ref{diff}). 

While outside the weak coupling regime the meaning of a coupling constant is ambiguous, this is not so for the order parameters. 
It is natural to conjecture that this monotonic behavior may be a universal property surviving to arbitrary coupling. 
As we move from the UV to the IR the WL order parameter decreases while the TL increases. We can explore a theory with adjoint matter such that it perturbs the exact YM theory but still confines in the IR. Or we might want to consider a theory with matter fine-tuned such that we lie in the conformal window. 
Reaching the IR, if there is confinement, the WL should vanish exponentially (area law) while the TL must then saturate to $\log|Z|$. Then, the monotonic behavior seems to be kept all the way down to the IR. At a conformal fixed point, both the WL and the TL get constant under scaling, and in the weak coupling regime are still described by the formulas we presented in section \ref{cert}. 

It is interesting to note that for a conformal fixed point the relative entropy $S_{WT}$, which takes into account both the WL and TL, has a fixed value at the special geometry of cross-ratio $\eta=1/2$. We have $S_{WT}=\log|Z|$ exactly because of the certainty relation \cite{Casini:2020rgj}. In the weak coupling regime, this is achieved mainly by the effect of the WL, while the subleading effect of the TL and the WL must exactly cancel to all orders for $\eta=1/2$. 

However, the monotonicity of order parameters is not universal. If we have a scenario of spontaneous symmetry breaking, even if the coupling constant might remain weak, our formulas for the optimal fluxes $\langle \Phi_{E,B}^2\rangle$ are modified by the mass of the vector fields. The optimal magnetic flux squared goes to zero above the SSB scale, while the electric one increases with the area \cite{Casini:2020rgj}. As a consequence, the WL relative entropy saturates asymptotically to $\log|Z|$ while the TL one goes to zero for large size with an area law. Therefore, the sign of the change with scale is opposite in the UV and IR in these models. The change of asymmetry between electric and magnetic parameters with scale switch direction in the middle of the flow.  This shows no universal monotonic behavior of these relative entropies is to be intrinsically expected, and further information on the dynamics and correlations is necessary to have predictive power about the IR fate of the symmetries based on their UV behavior.

\section*{Acknowledgements} 
We wish to thank Guillermo Silva for discussions and collaboration through the beginning of this project. This work was partially supported by CONICET, CNEA, and Universidad Nacional de Cuyo, Argentina. The work of H. C. is partially supported by an It From Qubit grant by the Simons Foundation. The work of J.M is supported by a DOE QuantISED grantDE-SC0020360 and the Simons Foundation It From Qubit collaboration (385592).

\appendix

 \section{Aspects of representation theory for Lie groups }\label{Lie}
\label{rept}

There are many excellent textbooks on the subject of Lie groups, both mathematics or physics-oriented, see \cite{brocker2003representations,Cornwell:1997ke,Costa:2012zz,hamermesh1989group,sternberg2009lie,
zee2016group,carter_macdonald_segal_taylor_1995,roman2011fundamentals}. The objective of this appendix is only to summarize certain topics in the theory of representations that have appeared in the main text, and that could take a while to grasp from the standard literature.

A Lie group $G$ is a group that has a manifold structure. Its group product and inverse are analytic functions in the manifold. It has an infinite number of elements and, more importantly, it can be analyzed near the identity. The tangent space of the Lie group at the identity is called the Lie algebra $\mathfrak{g}$. If a generic group element is specified by $d$ real parameters, the dimension of the tangent space is also $d$. This determines the number of independent generators $T_a$, $a=1,\cdots, d$  of the Lie algebra. These generators satisfy the usual commutation rules
\be 
[T_a,T_b]=f_{ab}^{c}T_c 
\ee
The numbers $f_{abc}$ are called the structure constants of the Lie algebra. They are antisymmetric in the first two indices and the Jacobi identity forces them to satisfy
\be 
f_{ab}^{c}f_{cd}^{e}+f_{bd}^{c}f_{ca}^{e}+f_{da}^{c}f_{cb}^{e}=0\;.
\ee
Understanding the classification of Lie algebras is tantamount to understand the space of solutions to the previous equation subject to the constraint $f_{ab}^{c}=-f_{ba}^{c}$. The works of Killing and Cartan led to such classification for simple Lie algebras, which are those containing no nontrivial invariant subalgebras. As it is well known, there are four infinite families and five exceptional ones. Since semisimple Lie algebras can be written as direct sums of simple ones, this classification extends to semisimple Lie algebras too.

The path towards such classification starts by noticing that the previous commutator can be seen as an action of $T_a$ on $T_b$. It defines an action of the algebra on itself. This is called the adjoint representation. By definition, it is a common representation for all Lie algebras. In coordinates, we have that
\be 
(T_a^{\textrm{adj}})_{bc}=f_{ab}^{c}\,.
\ee
This representation already allows to define a symmetric bilinear form, known as the Killing form 
\be \label{killing}
g_{ij}=-\textrm{Tr}_{\textrm{adj}} (T_i\, T_j )\;.
\ee
It can be diagonalized by a change of basis. By simplicity and convention, the basis of generators of the Lie algebra is always chosen to diagonalize this metric.\footnote{The Cartan criterion for semi-simplicity of a Lie algebra states that a Lie algebra is semi-simple if and only if the Killing form is non-degenerate. This implies that the adjoint representation for semi-simple Lie algebras is faithful. If the Lie algebra is simple, its adjoint representation is irreducible.} This metric will later allow us to identify the tangent space with its dual space.

More importantly, having a specific representation we can proceed to find the eigenvectors and eigenvalues of a maximal subalgebra of commuting generators. Since the vector space, in this case, is the Lie algebra itself, this provides us with a new basis for the Lie algebra. This maximal Abelian subalgebra is called the Cartan subalgebra $\mathfrak{h}$. Its number of generators $l$ is called the rank of the Lie algebra. The choice of Cartan subalgebra $\mathfrak{h}$ is of course not unique, but all choices are related by group transformations, as we comment further below when we discuss the equivalence classes in Lie groups.

Once such Cartan subalgebra is chosen, it is clear that they should be included in the basis of eigenvectors (with eigenvalue zero) of the adjoint representation since
\be 
[H_i,H_j]=0\,\,\,\,\, i=1,\cdots , l\,.
\ee
The second set of $d-l$ eigenvectors of the adjoint representation turns out to be even in number. They can be labelled as $E_{\pm \alpha}$, where $\alpha$ are $l$-components vectors, called the root vectors. These root vectors are the associated eigenvalues in the adjoint representation
\be\label{ca1}
[H_{i},E_{\pm\alpha}]=\pm\alpha_i \, E_{\pm\alpha}\;.
\ee
One can prove that these vectors are highly constrained from several perspectives. These properties or contraints play a fundamental role in the theory of representations and allow for the classification of simple Lie algebras. The remaining commutation relation can be proven to be
\bea \label{ca2}
&[E_{\alpha},E_{-\alpha}]=\sum\limits_{i=1}^{l}\alpha_i \, H_{i}& \\
& [E_{\alpha},E_{\beta} ]=N_{\alpha\beta} \, E_{\alpha+\beta}&\;,
\eea
In the second equation $N_{\alpha\beta}=N_{-\beta\alpha}\neq 0$ if and only if $\alpha+\beta$ is also a root vector.

Mathematically, root vectors are maps from the Cartan subalgebra $\mathfrak{h}$ to the real numbers. They are thus part of the dual space $\mathfrak{h}^{*}$. Indeed, root vectors are a basis of such dual space, if we allow arbitrary linear combinations. In this vein, it is interesting to check whether all of them are linearly independent. This is not the case and one can find a set of $l$ fundamental roots $\alpha^{(i)}$, with $i=1,\cdots ,l$ such that all other roots can be written as linear combinations with integer coefficients. This naturally suggests the definition of a lattice, the root lattice
\be 
\Lambda_{\alpha}\equiv \left\lbrace\sum\limits_{i=1}^{l} a_{i}\alpha^{(i)}\,\,\,\,\,\,\textrm{with}\,\,\,\,\, a_{i}\in \mathds{Z} \right\rbrace\,.
\ee
In terms of the representations of the theory, this lattice contains all the representations appearing in the decomposition in irreducible representations of arbitrary tensor products of the adjoint representation.

The Cartan subalgebra $\mathfrak{h}$ can be simultaneously diagonalized in every representation of the theory. In each representation $r$, we can find eigenvectors $\vert \psi_{a}^{r}\rangle$ and eigenvalues $\lambda_{i}^{ar}$, for $a=1,\cdots, d_r$, satisfying
\be 
H_i\vert \psi_{a}^{r}\rangle =\lambda_{i}^{ar}\vert\psi_{a}^{r}\rangle\;.
\ee
As for the roots, such $l$-component vectors $\lambda_{i}^{ar}$, with $i=1,\cdots ,l$, can be seen as maps from the Cartan subalgebra to the reals. These $l$-component vectors, one per eigenvector in each representation are called \emph{weights}. We see that the roots are the weights of the adjoint representation. Understanding the space of weights and their interplay with the full Lie algebra is tantamount to understanding the representations of the Lie algebra.

To understand the space of weights it is convenient to go back to the commutations relations of the Lie algebra, expressed in the form~(\ref{ca1}) and~(\ref{ca2}). The first relation resembles the commutation relation of $J_{z}$ with $J_{\pm}$ for the $SU(2)$ case. Indeed, for $SU(2)$ we have $l=1$, the weights and roots are numbers, the Cartan subalgebra is chosen to be $J_{z}$ by convention, and we have $J_{\pm}=E_{\pm 1}$. Luckily, the intuition from $SU(2)$ can be carried over to the generic case. We can define a bunch of $SU(2)$ subalgebras in the following manner
\bea 
J_{\pm}^{\alpha}&=&\frac{1}{\vert\alpha\vert}E_{\pm\alpha}\,\\
J_z^{\alpha}&=&\frac{1}{\vert\alpha\vert^2}\,\alpha\cdot H\;,
\eea
where $\alpha\cdot H =\sum\limits_{i}\alpha_i H_i$. Any irreducible representation is then going to be characterized by a highest weight $\Lambda$ of these subalgebras and a set of weights $\omega$ inside such irrep. Given the previous definitions, the associated eigenvector $\vert \Lambda,\omega\rangle$ is such that
\bea 
J_z^{\alpha}\vert \Lambda,\omega\rangle &=&\frac{\alpha\cdot\omega}{\vert\alpha\vert^2}\,\vert \Lambda,\omega\rangle\,,\\
J_{\pm}^{\alpha} \vert \Lambda,\omega\rangle &\propto &\vert \Lambda,\omega\pm\alpha\rangle
\;.
\eea
Since the eigenvalues of $J_z$ must be integers or half integers, we reach a fundamental conclusion. For all weights of a Lie group, it must be the case that
\be \label{dir1}
\frac{2 \alpha\cdot\omega}{\vert\alpha\vert^2}\in \mathds{Z}\;.
\ee
In particular, this applies to the roots themselves
\be \label{dir2}
\frac{2 \alpha\cdot\beta}{\vert\alpha\vert^2}\in \mathds{Z}\;,
\ee
which are the weights of the adjoint representation.\footnote{In fact the $SU(2)$ algebra structure implies more features. In particular it constraints the angle between roots to have some possible specific values. These restrictions play a key role in the classification of Lie algebras.}

As happens for the roots, we would want a set of fundamental weights, allowing us to reconstruct any weight by linear combinations with integer coefficients. It turns out that there are also $l$ fundamental weights $\omega^{(j)}$, with $j=1,\cdots , l$. These are defined by the following equations
\be 
\frac{2 \alpha^{(i)}\cdot\omega^{(j)}}{\vert\alpha^{(i)}\vert^2}=\delta_{ij}\;.
\ee
Intuitively, fundamental weights saturate the relation~(\ref{dir1}). It proves useful to give some different but equivalent definitions. The Killing form~(\ref{killing}) establishes a bijection between the Cartan subalgebra and its dual space. It allows us to define a new set of generators of the Cartan subalgebra $\tilde{H}_i$ such that
\be 
\alpha_j (\tilde{H}_i)=\frac{2 \alpha_i\cdot\alpha_j}{\vert\alpha_i\vert^2}\;.
\ee
The $\tilde{H}_i$ are called the fundamental co-roots. They are called co-roots because their coefficients in the Cartan basis, define a new set of dual roots $\alpha^{\vee}$, namely
\be 
\alpha^{\vee}_{(i)}\equiv \frac{2 \alpha^{(i)}}{\vert\alpha^{(i)}\vert^2}\;.
\ee
It can be verified that they indeed satisfy all defining properties of roots. Therefore they should characterize a dual Lie algebra. This dual Lie algebra, which has the same rank as the original one is called the Langlands dual, due to its role in the Langlands duality program \cite{10.1007/BFb0079065,Kapustin:2006pk}. In the physics community, the dual Lie algebra and associated dual Lie group are called electromagnetic or GNO dual group. This is due to the seminal work \cite{Goddard:1976qe}, where it was first noticed that the space of allowed monopoles satisfying the Dirac quantization condition is described by the representations of such group, as we further comment below. The fundamental co-roots also generate a lattice, called the co-root lattice $\Lambda_{\alpha^{\vee}}$. The physical meaning is dual to the root lattice meaning. Such lattice contains the weights of all representations appearing in the decomposition in irreducibles of arbitrary products of the adjoint representation of the dual Lie algebra.

Using the fundamental co-rrots, we can define the fundamental weights simply as
\be 
\omega^{(i)}(\tilde{H}_j)=\delta_{ij}\;.
\ee
Using the fundamental weights we can now write, in particular, the fundamental roots as
\be
\alpha^{(i)}=\sum\limits_{j=1}^{l} \frac{2 \alpha^{(i)}\cdot\alpha^{(j)}}{\vert\alpha^{(j)}\vert^2}\omega^{(j)}\equiv \sum\limits_{j=1}^{l} A_{ij}\,\omega^{(j)}\;,
\ee
where we have defined the so-called Cartan matrix
\be 
A_{ij}=\frac{2 \alpha^{(i)}\cdot\alpha^{(j)}}{\vert\alpha^{(j)}\vert^2}\;,
\ee
This is an $l\times l$ matrix of integer numbers.

As with the roots, the fundamental weights suggest a lattice, called the weight lattice
\be 
\Lambda_{\omega}\equiv \left\lbrace\sum\limits_{i=1}^{l} a_{i}\,\omega^{(i)}\,\,\,\,\,\,\textrm{with}\,\,\,\,\, a_{i}\in \mathds{Z} \right\rbrace\,.
\ee
The weight lattice contains all possible weights of the Lie group. In particular
\be \label{rinw}
\Lambda_{\alpha}\subset\Lambda_{\omega}\;.
\ee
The same (or dual) construction of fundamental ``magnetic'' weights can be done starting with the fundamental co-roots. The magnetic weights also generate a lattice $\Lambda_{\omega^{\vee}}$ and we have
\be 
\Lambda_{\alpha^{\vee}}\subset\Lambda_{\omega^{\vee}}\;.
\ee
It is a key result in the representation theory of Lie groups that any weight $\omega_{\textrm{dom}}$ that can be expressed as
\be
\omega_{\textrm{dom}}= \sum\limits_{i=1}^{l} a_i \,\omega^{(i)}\;,
\ee
where $a_{i}$ are non-negative integers, corresponds to one and only one irreducible representation. These are called ``dominant'' weights. Moreover, the previous dominant weight will appear when we fusion (tensor product of representations) $a_1$ times $\omega^{(1)}$, with  $a_2$ times $\omega^{(2)}$ and so on and so forth.

The weight lattice, as defined above, can be seen to emerge naturally and geometrically from the set of dominant weights by using the root vectors. This is the geometric counterpart of using the ladder operators $E_{\pm \alpha}$. An alternative way to move between different weights $\omega$ is through Weyl transformations
\be 
S_{\alpha} \omega=\omega-\frac{2 \alpha\cdot\omega}{\vert\alpha\vert^2}\alpha \;.
\ee
This can be seen to be a reflection of $\omega$ with respect to the plane perpendicular to $\alpha$. These transformations form a group, the Weyl group $W$. It can be proven that any weight $\omega\in\Lambda_{\omega}$ is related to a unique dominant weight $\omega_{\textrm{dom}}\in\Lambda_{\textrm{dom}}$ by a Weyl transformation. In other words, any weight is in the orbit of a unique dominant weight. Therefore we have the equivalence
\be 
\Lambda_{\textrm{dom}}\sim \Lambda_{\omega}/W\;.
\ee
The Weyl group leaves invariant the inner product between weights.

The weight and root lattice just defined are useful for understanding the rules for decomposing tensor products of representations. There are several geometric rules in this context which we will not review here. But we want to make a final remark related to this. The weight lattice contains all weights of the Lie group. The root lattice, in contrast, contains only those weights associated with irreducible representations appearing in tensor products of the adjoint representation. It is transparent that one is contained in the other~(\ref{rinw}). It is thus natural to ask for the quotient, which is known to be\footnote{Here we are implicitly assuming that we are working with the universal covering group associated with the given Lie algebra. Therefore we have all possible weights, as it is clear when we define the weight lattice using all the fundamental weights. The same applies to the statements for the dual group. Several intermediate choices appear as well, see \cite{Kapustin_2006,Aharony:2013hda}. From our perspective, those choices pertain more to a given definition of a net of algebras, as described in \cite{Casini:2020rgj}, and not the analysis of the full set of operators of the theory.}
\be 
\Lambda_{\omega}/\Lambda_{\alpha}\sim Z^*\sim Z\;,
\ee
where $Z^*$ is the group of irreducible representations of the center of the gauge group. This is a finite abelian group and therefore isomorphic by Pontiagryn duality to the center itself. Since the root lattice contains all representations appearing in products of the adjoint, the previous quotient tells us the classes of non-additive loops, the ones that have been the focus of the article and that are labeled by elements of the center of the group. A related observation that can be derived from these results is that while a weight (a point in the weight lattice) is not associated with only one irreducible representation of the gauge group $G$, it is associated with one and only one irreducible representation of $Z_{G}^*$. A similar result holds for the dual lattices
\be 
\Lambda_{\omega^{\vee}}/\Lambda_{\alpha^{\vee}}\sim  Z \sim Z^* \;.
\ee

\section{Maxwell orbifold}
\label{weak}

At weak coupling, the non-Abelian gauge theory reduces dynamically to that of independent free Maxwell fields
\be
F_{\mu\nu}^a=\partial_\mu A^a_\nu-\partial_\nu A^a_\mu + g f^{abc} A_\mu^b A_\nu^c \rightarrow \partial_\mu A^a_\nu-\partial_\nu A^a_\mu \;.
\ee
In this simplified Abelian description, we have to be careful about the allowed gauge transformations that respect the limit. For a non-Abelian gauge field, we have the gauge transformation
\be
A_\mu'=U\, A_\mu \, U^\dagger -\frac{i}{g} \, (\partial_\mu U)\, U^\dagger \,.\label{haha}
\ee
 To take the limit (\ref{ff}) we assumed $A\sim g^0$. To keep this order we are not allowed to make arbitrary gauge transformations. Eq. (\ref{haha}) shows we have to take a slowly varying $U$ with derivatives which are of order $g$. We then write
\be
U(x)=U_0 \, e^{i g \phi(x)}\,,  
\ee 
with $\phi(x)$ in the Lie algebra and order $g^0$ and $U_0$ constant. The gauge transformation is 
\be
A'_\mu=U_0(A_\mu+ \partial_\mu \phi)U_0^\dagger\,.
\ee
It is composed of two independent transformations. One is the gauge transformation of $d_G$ Maxwell fields
\be
A_\mu^{a '}=A_\mu^a+\partial_\mu \phi^a\,,
\ee
and the other is a global rotation. These are the symmetries of the limit theory.  Then the limit theory is one of independent Maxwell fields but where we have to consider only operators invariant under global transformations of the group. In other words, it is an orbifold of $d_G$ Maxwell fields, with $d_G$ the dimension of the Lie algebra.

Let us study the structure of the Maxwell orbifold algebras for a ring. In the ring, the additive operators are $G$ invariant combinations of the electromagnetic curvature tensor with coordinates inside $R$. Non-local operators should involve non-local operators for each of the Maxwell fields. A generating basis for these is formed by the magnetic and electric fluxes $\Phi^a_B,\Phi^a_E$. In the orbifold, we have to take group invariant combinations,  like the quadratic invariant $\Phi_B^a \Phi_B^a$, and more generally the Casimir invariants 
\be 
c_{a_1 \cdot a_n} \Phi_B^{a_1}\cdots \Phi_B^{a_n}\,,
\ee
where the tensor $c$ is symmetric and gives place to an order $n$-Casimir when the fluxes are replaced by Lie algebra elements. There are as many algebraically independent Casimirs as the rank of the Lie algebra. The non-local operators are generated by arbitrary polynomials in these electric and magnetic invariants, and there can be mixed electric and magnetic invariants too. We can organize a generating set of non-local operators in different ways. We have for example the smeared loops
\be    
W_{r,q}= \tr (e^{i \, q\, \Phi_B^r})\,,\label{tyy}
\ee
for all charges and representations. Here $\Phi_B^r =\Phi_B^a T^r_a$.  There are their electric counterparts too. A series expansion of these loops is composed of the polynomial invariant described above. We can also rewrite these invariants in a 't Hooft loop form
\be
T_{c,q}=E_G( e^{\frac{i}{q} \,\lambda_ a\, \Phi_E^a  })\,,\label{tyu}
\ee
where $c$ is a label to a conjugacy class on $G$ that depends on $\lambda$ and we describe shortly, and $E_G$ is the average over the global group rotations. There are the magnetic counterparts too. 
These do not involve the Lie algebra matrices of the representations. Expanding in series we find again invariant polynomials.  
Outside the ring the exponentials $e^{\frac{i}{q} \,\lambda_ a\, \Phi_E^a}$ act introducing a translation into the magnetic fluxes. We have
\be
e^{\frac{i}{q} \,\lambda_ a\, \Phi_E^a}\, (q\,\Phi^b_B)\, e^{-\frac{i}{q} \,\lambda_ a\, \Phi_E^a}= q\,\Phi_B^b + \,\lambda^b\,.
\ee
So these exponentials can be assimilated to translations in the Lie Algebra by the element $\lambda^b T_b$, associated to a group element $e^{i\,\lambda^b T_b}$. The conditional expectation in (\ref{tyu}) is then naturally associated with the conjugacy class $c$ of this group element.  

In both representations (\ref{tyy}), (\ref{tyu}), we have non-Abelian fusion rules of representations and conjugacy classes respectively.\footnote{This does not contradict the theorem in \cite{Casini:2020rgj} stating that commutation relations for non-local operators in rings in $d=4$ correspond to Abelian groups. That theorem holds when there are no non-local sectors for two balls. This is certainly not the case for an orbifold like the present one.} Their commutation relations are complicated to describe, see the analogous case for non-Abelian global symmetries in \cite{Casini:2020rgj}. 

A natural question is whether the limit of the relative entropy of the order parameters for the non-Abelian theory corresponds to some specific relative entropy on the orbifolded Maxwell theory in the weak coupling limit. That is, we have to understand if the limit of the algebras and conditional expectations correspond to some structures in the orbifold. This is a question we cannot answer in the negative, but our failed attempts at this identification lead us to be suspicious about a positive answer. On one side we could not identify, among the infinitely many classes of non-local operators on the ring, some particular non-local operators having the commutation relations expected for the non-local operators in the non-Abelian model. In particular, smeared magnetic loops labeled by representations and TL (\ref{tyu}) labeled by the center do not commute as they ought to do.   

This is perhaps not that surprising because of the following reason. The first point is that all operators in the Maxwell orbifold do correspond to limits of operators in the non-Abelian theory. In fact, the generators of the orbifold are $G$ invariant combinations of $F^a_{\mu\nu}$ fields, such as $F^a_{\mu\nu}(x)F^a_{\alpha\beta}(y)$. This operator can be thought of as a limit of Wilson lines in the non-Abelian model. This shows one of the problems that appear when identifying algebras between the two models. The relation between operators is many to one since, for example,  different Wilson lines with the same ends go to the same operator. Another delicate problem appears in the identification of algebras and regions. The operator  $F^a_{\mu\nu}(x)F^a_{\alpha\beta}(y)$ is a non-local operator of two balls in the orbifold theory, while it cannot be considered an operator of two balls in the full theory.

There is another, perhaps more severe, difficulty in this attempted identification.  Very few non-local operators in the ring in the orbifold could come from non-local operators in the ring in the full theory.  
There are local operators such as adjoint WL that are breakable in the full theory but cannot be broken in the orbifold. We might take this into account by considering the limit of the additive algebra of the full theory as the true additive algebra of the orbifold.  Then, there are also non allowed charges in the non-local operators of the orbifold. These can be thought of as particular combinations of magnetic fluxes in the Maxwell theory, and probably can be reconstructed as limits of gauge invariant surface operators in the full theory. These combinations must not be considered part of the algebra of the ring, but as surface operators instead.

\bibliographystyle{JHEP}
\bibliography{EE}

\end{document}